\documentclass{aa}
\usepackage{graphicx}
\usepackage{txfonts}
\usepackage[]{natbib}
\begin{document}
   \title{
   X-ray flares on the \mbox{UV Ceti-type} star \mbox{CC Eridani}:\\
   a ``peculiar'' time-evolution of spectral parameters\thanks{Based 
   on observations obtained with XMM-Newton, an ESA science mission 
   with instruments and contributions directly funded by ESA Member 
   States and NASA.}}


   \titlerunning{XMM-Newton observation of the flare star \mbox{CC Eri}}

   \author{I. Crespo-Chac\'on
          \inst{1,2}
          \and
          G. Micela
          \inst{1}
          \and
          F. Reale
          \inst{1,3}
          \and
          M. Caramazza
          \inst{1,3}
          \and
          J. L\'opez-Santiago\thanks{Presently at Departamento de Astrof\'{\i}sica, Facultad de Ciencias
              F\'{\i}sicas, Universidad Complutense de Madrid, E-28040
              Madrid, Spain (jls@astrax.fis.ucm.es).}
          \inst{1}
          \and
          I. Pillitteri
          \inst{3}
          }

   \offprints{I. Crespo-Chac\'on}

   \institute{INAF - Osservatorio Astronomico di Palermo, Piazza del Parlamento 1, I-90134 Palermo, Italy\\
              \email{ichacon@astropa.unipa.it, giusi@astropa.unipa.it, jlopez@astropa.unipa.it}
         \and
              Departamento de Astrof\'{\i}sica, Facultad de Ciencias
              F\'{\i}sicas, Universidad Complutense de Madrid, E-28040 Madrid, Spain\\
             \email{icc@astrax.fis.ucm.es}
         \and
              Dip. di Scienze Fisiche e Astronomiche -- Sez. di Astronomia -- Universit\`a di Palermo, Piazza del Parlamento 1, I-90134 Palermo, Italy\\
             \email{reale@astropa.unipa.it, mcarama@astropa.unipa.it, pilli@astropa.unipa.it} 
             }

   \date{Received 4 April 2007 / Accepted 7 June 2007}


  \abstract
   {
   Weak flares are supposed to be an important heating agent of the 
   outer layers of stellar atmospheres. However, due to instrumental 
   limitations, only large X-ray flares have been studied 
   in detail until now.
   }
   {
   We used an \mbox{XMM-Newton} observation of the very active
   \mbox{BY-Dra} type binary star CC Eri in order to investigate 
   the properties of two flares that are weaker than those typically 
   studied in the literature.
   }
   {
   We performed \mbox{time-resolved} spectroscopy of the data taken with 
   the \mbox{EPIC-PN} CCD camera. A \mbox{multi-temperature} 
   model was used to fit the spectra.
   We inferred the size of the flaring loops
   using the \mbox{density-temperature} diagram.
   The loop scaling laws were 
   applied for deriving physical parameters of the flaring plasma. 
   We also estimated the number of loops involved in
   the observed flares.
   }
   {
   A large X-ray variability was found.
   Spectral analysis showed that all the regions in the light curve,
   including the flare segments, are \mbox{well-described} by a \mbox{3-$T$}
   model with variable emission measures but, surprisingly, with constant temperatures
   (values of 3, 10 and 22 MK). The analysed flares
   lasted \mbox{$\sim$ 3.4} and 7.1 ks, with flux increases 
   of factors \mbox{1.5 -- 1.9}.
   They occurred in arcades made of a few tens of similar coronal loops.
   The size of the flaring loops is much smaller than the distance between the stellar
   surfaces in the binary system, and even smaller than the
   radius of each of the stars.
   The obtained results are consistent with the following ideas: (i) the whole X-ray light
   curve of CC Eri could be the result of a superposition of multiple
   \mbox{low-energy} flares, and (ii) stellar flares can be
   \mbox{scaled-up} versions of solar flares.
   }
   {}


   \maketitle
%

\section{Introduction}
\label{introduction}

Solar-like stars (main sequence stars with spectral types from F to early M)
have radiative cores and convective outer envelopes. Convection, together
with differential rotation, generates a magnetic dynamo
\citep[see][and references therein]{1975ApJ...198..205P} that is 
responsible of the formation of the corona
\citep[see][for two extensive reviews about stellar coronal astronomy]{2003SSRv..108..577F,2004A&ARv..12...71G}. 
Magnetic activity similar
to that observed on the Sun is typically detected on these stars,
showing variability through all the electromagnetic spectrum.
The activity level of a star is frequently measured in terms of
its coronal \mbox{X-ray} luminosity $L_\mathrm{X}$.
Using data collected by the ROSAT satellite,
\citet{1995ApJ...450..392S} and \citet{1997A&A...318..215S} found 
that solar-like stars present \mbox{X-ray} luminosities 
(integrated over the \mbox{0.1-2.4 keV} energy band) in the range 
\mbox{25.5 $\lesssim \mathrm{log}~L_\mathrm{X} (\mathrm{erg~s^{-1}}) \lesssim$ 29.5}.
This luminosity is correlated with the stellar rotation rate
\citep{1981ApJ...248..279P}. \citet{1982ApJ...253..745W} noticed
that a single power law dependence between $L_\mathrm{X}/L_\mathrm{bol}$ and
the angular velocity was unable to reproduce all the
observed data, and proposed to replace it by either a
broken power law or an exponential relationship.
This result later led to the concept of saturation of stellar
activity at high rotation rates 
\citep{1984A&A...133..117V,1987ApJ...321..958V}.
Rotational velocity, and therefore activity, decreases
with age because of the angular momentum losses due to
the magnetised stellar wind \citep{1984ApJ...283L..63R,2004A&A...426.1021P}. 
However, \mbox{short-period}
binaries can maintain high rotation rates since they are
synchronized by tidal coupling.

Flares are the most extreme evidence of magnetic activity
in stellar atmospheres. Frequent flaring is found on late
K and M dwarfs in the solar neighbourhood (the so-called
UV \mbox{Ceti-type} stars -- see their general properties
in Pettersen, 1991). 
Similarities between solar
flares and those observed on UV \mbox{Ceti-type} stars suggest that
they are produced by the same basic physical mechanisms.
Flares are supposed to be the result of the energy release from
magnetic field reconnection in the lower corona \citep[e.g.][]{1976SoPh...50...85K}. 
In consequence, electrons and ions are accelerated and gyrate
downward along the magnetic field lines, producing
synchrotron radio emission. Bremsstrahlung
radiation is emitted in hard X-rays (\mbox{$>$ 20 keV}) when these
ionized beams collide with the denser material of
the chromosphere. At the same time, the gas in the
affected chromospheric region is heated
(optical and UV radiation is then emitted)
and evaporated. Thus, the density and
temperature of the newly formed coronal loops increase,
emitting in soft X-ray and extreme UV wavelengths.
Recent evidence in favour of this scenario was
given by \citet{2005A&A...431..679M} and \citet{2005A&A...436..241S}. 

A recurrent idea in the literature is that flares
are the main heating agent of the outer stellar
atmospheres \citep[e.g.][]{2000ApJ...541..396A}, 
so that the observed ``quiescent'' emission
would be the result of a superposition of multiple
small flares (named \mbox{``nano-flares''}). 
The distribution law that gives the number of
flares $dN$ within an \mbox{energy-interval} $[E,E+dE]$
allows us to estimate the energy budget of the
corona. In particular, solar flares are 
distributed following a power law \citep{1974SoPh...39..155D,1984ApJ...283..421L,1985SoPh..100..465D},
that is
\begin{equation}
  \frac{dN}{dE} = k_1 E^{-\alpha},
 \label{eq:flaresdistribution}
\end{equation}
where $k_1$ is a constant and $\alpha$ is the \mbox{power-law} index. 
For $\alpha > 2$, an extrapolation of \mbox{Eq.~(\ref{eq:flaresdistribution})}
to flare energies below the detection threshold would
be sufficient to account for the luminosity of the
quiescent corona. For this reason, it is crucial
to investigate the validity of such an extrapolation \citep{1991SoPh..133..357H}, as
well as the value of $\alpha$ on magnetically active stars.
For the Sun, values of $\alpha$ between 1.5 and 2.6  
were reported \citep{1993SoPh..143..275C,1995ApJ...438..472P,1998ApJ...501L.213K}.
However, a later study carried out by \citet{2000ApJ...535.1047A} 
suggests the insufficiency of \mbox{nano-flares}
to heat
the solar corona. 
On the other hand, for active stars, 
\citet{1988A&A...205..197C} and \citet{1999ApJ...515..746O} 
found $\alpha \approx 1.5-1.6$, while other
authors have reported $\alpha > 2$
\citep{1999ApJ...513L..53A,2000ApJ...541..396A,2002ApJ...580.1118K,2003ApJ...582..423G}.
Therefore, it is still unclear if \mbox{Eq.~(\ref{eq:flaresdistribution})}
can be extrapolated from large observable flares towards the weakest ones
in order to support the \mbox{flare-heating} hypothesis.

While flare stars have been studied in the optical
during more than half a century, the first
sizeable sample of \mbox{X-ray} flares was only compiled
after the launch of the {\it Einstein} Observatory
\citep{1983ards.proc..255H}. Later,
\citet{1990A&A...228..403P} presented the results
of a comprehensive survey of \mbox{X-ray} observations
of flare stars carried out with the EXOSAT Observatory.
However, until recent years, and due to instrumental 
limitations, only large \mbox{X-ray} flares
could be studied in detail. 
At the present time, the great sensitivity, wide energy
range, high energy resolution, and continuous time coverage
of the EPIC (European Photon Imaging Cameras)
\mbox{detectors -- on-board} the \mbox{XMM-Newton} \mbox{satellite -- also} enable the
detection and analysis of smaller flares. UV \mbox{Ceti-type} flare
stars are specially indicated for this purpose because of its
proximity. For all these reasons, we decided to carry out the present 
study of the 
\mbox{X-ray} flares detected 
on the UV \mbox{Ceti-type} star
\mbox{CC Eri} using the \mbox{XMM-Newton} satellite. 

\mbox{CC Eri} (HD 16157) is a spectroscopic binary star (BY \mbox{Dra-type})
located in the immediate solar neighbourhood, at a distance of \mbox{11.51 $\pm$ 0.11 pc}
\citep[from Hipparcos,][]{1997A&A...323L..49P}.
This is a SB2 system \citep{1993A&AS..100..173S} which consists of a K7.5Ve primary and a M3.5Ve secondary 
\citep{2000A&A...359..159A}, with mass ratio $\approx 2$
\citep{1959MNRAS.119..526E,2000A&A...359..159A}. 
The photometric period \mbox{-- 1.56 days --} results to be
equal to that of the orbital motion \citep{1959MNRAS.119..526E,1973MNRAS.164..343B,1977AJ.....82..490B}.
Thus, the synchronization due to the tidal lock makes the primary component
to be one of the fastest rotating late K dwarfs in the solar vicinity.
Using kinematical criteria, the age of the system was estimated to be \mbox{9.16 Gyr}
\citep{2006MNRAS.366.1511D}.
\citet{1977A&A....60L..27B} and \citet{2000A&A...359..159A} found that 
the chromospheric emission of \mbox{CC Eri} varies in antiphase 
with its optical continuum, suggesting the presence of
active emission regions associated with starspots.
Besides, its quiescent radio emission is polarized at the 10 - \mbox{20~\%} level 
\citep{2002ApJS..138...99O,2004PASA...21...72S}, indicating \mbox{large-scale} ordering in the stellar magnetic field.
\mbox{CC Eri} presents a strong flare activity over a wide
range of energies 
\citep{1976IBVS.1186....1B,1978A&A....64..153B,1988AJ.....95..887C,1992LNP...397..255B,1992A&A...264L..31G,1995MNRAS.272...11P,2000A&A...359..159A,2002ApJS..138...99O,2004PASA...21...72S}. 
First \mbox{X-ray} detections of \mbox{CC Eri} were done with 
HEAO1, showing $\mathrm{log}~L_\mathrm{X} (\mathrm{erg~s^{-1}}) \approx 29.26$
in the \mbox{2 -- 20 keV} band \citep{1982ApJ...262..263T},  
{\it Einstein}, which measured $\mathrm{log}~L_\mathrm{X} (\mathrm{erg~s^{-1}}) \approx 29.51$
in the \mbox{0.15 -- 4.5 keV} range \citep{1982AJ.....87..558C}, and
EXOSAT, that observed $\mathrm{log}~L_\mathrm{X} (\mathrm{erg~s^{-1}}) \approx 29.62$ 
in the \mbox{0.04 -- 2 keV} interval \citep{1988A&A...191..109P}.
However, since none of these observations lasted very long,
little information on the temporal and spectral
variation of the source in \mbox{X-rays} was available.
\citet{1995MNRAS.272...11P} observed and analysed 
for the first time an \mbox{X-ray} flare on \mbox{CC Eri}, using 
ROSAT observations. They measured a quiescent luminosity 
$\mathrm{log}~L_\mathrm{X} (\mathrm{erg~s^{-1}}) \approx 29.40$
in the \mbox{0.17 -- 2 keV} band. The flare had an \mbox{e-folding}
rise and decay times of about \mbox{1 h} (or less) and \mbox{2 h},
respectively, 
and the emission was enhanced by a factor greater \mbox{than 2}. A \mbox{2-$T$}
model gave an adequate description to all the ROSAT spectra
and showed the presence of \mbox{high-temperature} plasma 
\mbox{($\sim 10$ MK)} even during the
\mbox{time-intervals} where no flare activity was detected.
This is consistent with the results obtained
for the quiescent emission of M-dwarfs 
using data from the EXOSAT ME and 
{\it Einstein} IPC \citep[e.g.][]{1990A&A...228..403P,1990ApJ...365..704S}.
All the \mbox{X-ray}
luminosities measured for \mbox{CC Eri} are similar 
and place this binary star among the most active ones.
This is clearly noticed when comparing the quiescent 
\mbox{X-ray} luminosity obtained for \mbox{CC Eri}
from the data analysed in this work
(\mbox{$L_\mathrm{X,~0.1 - 2.4~keV} \approx 3.7 \times 10^{29}$ $\mathrm{erg~s^{-1}}$}) 
with the cumulative \mbox{X-ray} 
luminosity distribution functions found by \citet{1995ApJ...450..392S} 
for \mbox{low-mass} stars in the solar neighbourhood.

In this paper, we present the study of an \mbox{XMM-Newton} observation
of \mbox{CC Eri}, which, for the first time, reveals details about the behaviour of the
parameters that characterized the plasma during two flares
weaker than those typically analysed in other active dM stars. 
Technical information
about the observation and details of the data analysis is given in \mbox{Sect.~\ref{sec:observations}}.
In \mbox{Sect.~\ref{sec:lightcurve}} we describe the light curve
and the different kinds of variations observed. 
The time-resolved study of spectral parameters
is presented in \mbox{Sect.~\ref{sec:spectra}},
where we also estimate the size of the flaring loops.
Finally, in \mbox{Sect.~\ref{sec:discussionconclusions}},
we discuss and interpret the results in the context of solar and stellar flares.


\section{Observations and data analysis}
\label{sec:observations}

   \begin{figure}
   \centering
      \includegraphics[bb=40 571 346 808,width=8.6cm,clip]{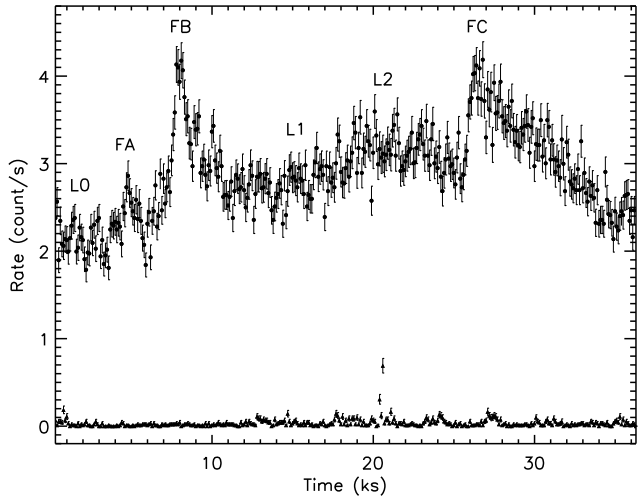}
      \caption{Source
      (upper region) and background
      (lower region)
      light curves of \mbox{CC Eri} as observed with the
      EPIC-PN detector for the energy band \mbox{0.5 -- 10.0 keV}
      and a temporal binning of 100~s.
      The labels FA, FB and FC refer to the
      detected flares while L0, L1 and L2 designate
      different emission levels where no clear
      flare activity was observed.
              }
         \label{fig:cceri_pn_lightcurve}
   \end{figure}
%

The observation of \mbox{CC Eri} analysed in this work was
performed with the \mbox{XMM-Newton} satellite on August, 8, 2003 (\mbox{PI:
H. Kay}, \mbox{ID: 0148790101}) during revolution 0671.
The \mbox{XMM-Newton} satellite owns the most sensitive soft 
\mbox{X-ray} detector system presently available:
the EPIC instrument.
It
consists of three imaging and non-dispersive \mbox{CCD-based} cameras:
the twin \mbox{MOS 1} and \mbox{MOS 2}, and the PN
\citep{2001A&A...365L..27T,2001A&A...365L..18S}.
The wavelength range measured by the EPIC detectors allows one
to obtain a reliable determination of properties of the hottest 
plasma components. 
For our timing and spectral analysis, we used only the data from the \mbox{PN-CCD}
camera, that is more sensitive than the MOS detectors.
This observation was done in the full frame mode with 
the thick filter.
The exposure time of the image taken
with \mbox{EPIC-PN} was \mbox{36.703 ks}, 
that is,
0.27 times the orbital period of the binary star \mbox{CC Eri}.

We used the event file in the PPS data products,
which was produced with the standard XMM-Newton Science Analysis
System (SAS) software, \mbox{version 5.4.2}.
The light curve and the spectra were obtained with
standard tools of SAS. 
The PN responses were generated with the SAS {\small RMFGEN} and
{\small ARFGEN} tasks.
The spectral analysis was done with the X-ray spectral fitting package
XSPEC V11.3.2 \citep{1996ASPC..101...17A,2004HEAD....8.1629A}.

Standard selection criteria were applied for
filtering the data \citep[see][]{2004UGXMMSAS}. 
We extracted events in the energy band between
\mbox{0.5 -- 10.0 keV} that triggered only 
one or two detector pixels at the same time (\mbox{PATTERN $\le$ 4}).
Data below 0.5~keV were excluded to avoid 
residual calibration problems
in the response matrices at soft energies. Nevertheless,
since flares mainly affect hotter thermal components,
the spectral region below 0.5~keV is not
crucial for our analysis.
In fact, we checked that no significant variations
are found in the spectral parameters
(see Sect.~\ref{subsec:spectral_parameters}) 
when the region \mbox{0.3 -- 0.5 keV} is included for
fitting the spectra.
The {\small EPATPLOT} task was used for confirming 
the existence of \mbox{pile-up} affecting the inner region of \mbox{CC Eri}.
To lose the minimum number of counts as possible,
we looked for and ignored the smallest region that allowed avoiding
the \mbox{pile-up} effects during the whole observation.
The X-ray light curve and spectra were therefore obtained
with the events taken from an annulus with inner radius 12\arcsec~and
outer radius 44\arcsec. On the other hand,
background photons were extracted
from a source-free region \mbox{-- a} circle with a radius
\mbox{of 54.4\arcsec --} placed at the same CCD as \mbox{CC Eri}.


%
\begin{table}
\caption{Time-intervals of the main activity levels
         detected on \mbox{CC Eri}~(see Fig.~\ref{fig:cceri_pn_lightcurve}).}
\label{tab:selectedperiods}
\centering
\begin{tabular}{c c}
\hline\hline
\noalign{\smallskip}
   Activity level & Time-interval (ks after the observation starts) \\
\noalign{\smallskip}
\hline
\noalign{\smallskip}
   L0 & ~0.0 -- ~3.6 \\
   FA & ~3.6 -- ~5.6 \\
   FB & ~5.6 -- 12.9 \\
   L1 & 12.9 -- 17.7 \\
   L2 & 17.7 -- 23.7 \\
   FC & 23.7 -- 36.7 \\
\noalign{\smallskip}
\hline
\end{tabular}
\end{table}
%

%
\begin{table}
\caption{Duration of the rising ($\tau_{\rm R}$) 
         and decay ($\tau_{\rm D}$) phases of the
         flares observed on \mbox{CC Eri}.}
\label{tab:timeflarephases}
\centering
\begin{tabular}{c c c}
\hline\hline
\noalign{\smallskip}
   Flare & $\tau_{\rm R}$ (s) & $\tau_{\rm D}$ (s)\\
\noalign{\smallskip}
\hline
\noalign{\smallskip}
   FA & 400  & 580  \\ 
   FB & 570  & 2850 \\ 
   FC & 1100 & 5960 \\ 
\noalign{\smallskip}
\hline
\end{tabular}
\end{table}
%

\section{The light curve}
\label{sec:lightcurve}

   \begin{figure*}
   \centering
      \includegraphics[bb=30 18 683 188,width=18cm,clip]{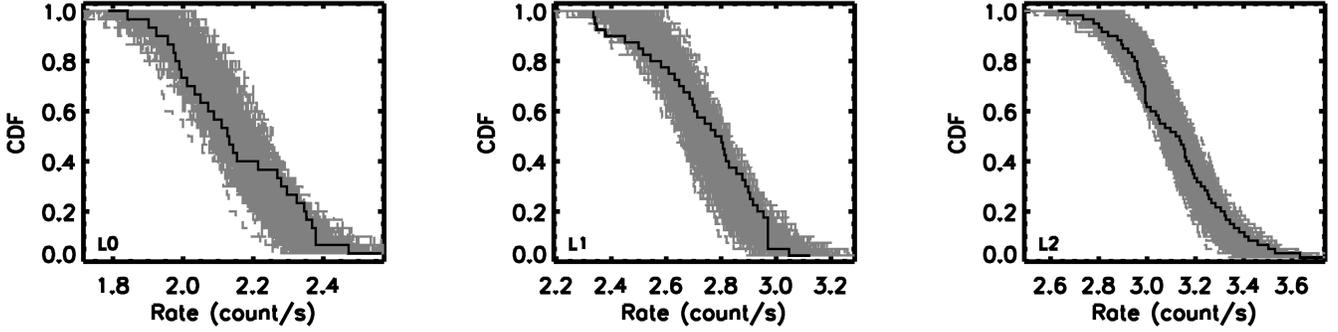}
      \caption{Cumulative distribution function of the \mbox{count-rates} observed
               in the regions L0, L1 and L2 (solid black line) 
               compared to those simulated by supposing a constant source
               with \mbox{count-rate} equal to the mean value
               ($R_\mathrm{m}[\mathrm{Li}]$) in the given region (dashed area).
              }
         \label{fig:poissonLOL1L2}
   \end{figure*}
%

   \begin{figure}
   \centering
      \includegraphics[bb=33 16 342 266,width=8.6cm,height=7.1cm,clip]{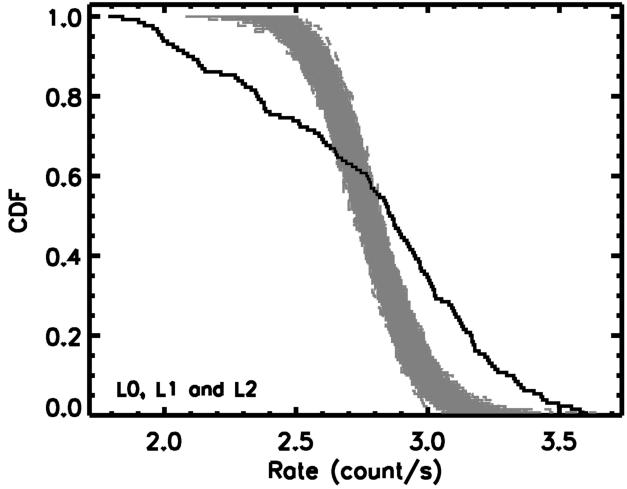}
      \caption{Cumulative distribution function of the observed \mbox{count-rates}
               obtained for L0, L1 and L2 altogether (solid black line) 
               compared to those simulated by supposing a constant source
               with \mbox{count-rate} equal to the mean value of these
               three regions (dashed area). 
              }
         \label{fig:poissonL0L1L2altogether}
   \end{figure}
%

   \begin{figure}
   \centering
      \includegraphics[bb=33 16 342 266,width=8.6cm,height=7.1cm,clip]{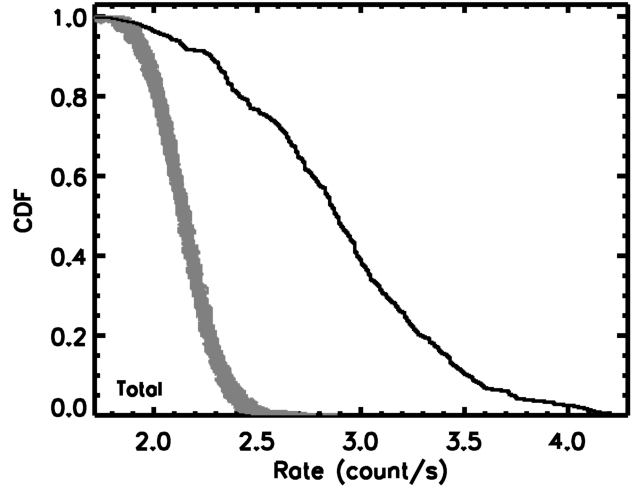}
      \caption{Cumulative distribution function of the \mbox{count-rates}
               detected in the whole observation of \mbox{CC Eri} 
               (solid black line) compared to those simulated by 
               supposing a constant source with \mbox{count-rate} equal to 
               the quiescent value (dashed area).
              }
         \label{fig:poissontotal}
   \end{figure}
%

Fig.~\ref{fig:cceri_pn_lightcurve} shows the source \mbox{(background - subtracted)}
and background light curves of \mbox{CC Eri}
as observed with the EPIC-PN detector for
a temporal binning of 100~s and the filtering criteria
given in Sect.~\ref{sec:observations}.
Note that the background \mbox{count-rate} was
scaled to the size of the source region.
The dead-time correction was applied to both
the source and background \mbox{count-rates}.
A significant \mbox{X-ray} variability was found
throughout the observation. The lowest activity level 
(L0 or {\it quiescent state} hereafter) was observed at the 
beginning. It was followed by a small flare (FA) where the
stellar flux increased by a factor 1.3. Another two flares (FB and FC)
with flux changes of factors \mbox{1.5 -- 1.9}, depending on the
selected reference level, were also observed. The
maxima of these two flares were separated by \mbox{5 hours}.
In addition, the stellar emission between these two events was
also variable \mbox{-- two}
different activity levels (L1 and L2) were \mbox{identified --}
and higher than that observed during the quiescent state.
The start and end times of the main selected periods are
summarized in Table~\ref{tab:selectedperiods}.
FA, FB and FC radiated in the \mbox{0.5 -- 10.0 keV} band
a total energy of 0.08,
0.75 
and \mbox{1.5~$\times~10^{33}$ erg}, respectively.

In terms of relative increasing of flux with respect to the
quiescent level,
the strength of the detected flares is somewhat smaller
than the typical values analyzed on active low-mass stars.
For instance, \citet{2005A&A...435.1073R} found
flux increases of factors \mbox{2 -- 3} in flares produced by
\mbox{EQ Peg}, \mbox{AT Mic}, \mbox{AD Leo} and \mbox{EV Lac}.
All of them can be considered as moderate flares
in view of the flux increases (\mbox{10 -- 300} times the quiescent
state value) that
\citet{2000A&A...353..987F}, \citet{2002ASPC..277..515K}, and 
\citet{2004A&A...416..713G}
observed respectively on \mbox{EV Lac},  \mbox{EQ Peg} and \mbox{Prox
Cen}.
The detected flares are even a little bit weaker than that observed by
\citet{1995MNRAS.272...11P} on the same star, \mbox{CC Eri},
which showed a ratio between the peak and minimum fluxes
larger than a factor \mbox{of 2}.
However, in absolute terms, the peak X-ray luminosity of the analysed 
flares is large compared to that of the brightest solar flares.

Table~\ref{tab:timeflarephases} lists the duration of 
the rising and decay phases of the observed flares.
Since the behaviour of these two phases can be 
\mbox{well-described} by an exponential law, 
their duration was respectively
estimated as the \mbox{1/{\it e}} rise time ($\tau_{\rm R}$) 
or \mbox{1/{\it e}} decay time ($\tau_{\rm D}$).
The values of $\tau_{\rm R}$ and $\tau_{\rm D}$ were therefore 
determined from a \mbox{least-squares} fit to the 
corresponding data by an exponential function of the form
\mbox{$R = A_0 e^{(t-t_\mathrm{max})/\tau} + R_\mathrm{base}$};
where $R$ is the \mbox{count-rate}, $R_\mathrm{base}$
is the \mbox{count-rate} in the quiescent state (L0 in this case),
$A_0$ is the amplitude at the flare maximum, $t$ is the time, $t_\mathrm{max}$
is the time at the flare maximum, and
$\tau$ is $\tau_{\rm R}$ \mbox{or - $\tau_{\rm D}$}.
Note in Fig.~\ref{fig:cceri_pn_lightcurve} that flare FC could 
have a larger or shorter decay time, depending on the
value considered for $R_\mathrm{base}$
(i.e., equal to that of L0 or, on the contrary, L1),
but being always greater than 50 minutes.
Using data collected by the LE experiment on EXOSAT,
\citet{1990A&A...228..403P} found two
different types of flares in a sample of 32 M dwarf stars,
i.e., {\it impulsive flares} and {\it long decay flares},
similar to the ones observed on the Sun \citep{1977ApJ...216..108P}.
Those in the first group are reminiscent of solar {\it compact} flares,
showing rise times of a few minutes
and decay times of tens of minutes; and those in the second group,
with decay times of the order of \mbox{$\approx$ 1 hour}
or longer, are reminiscent of solar  \mbox{long-duration}
{\it \mbox{2-ribbon}} flares. Thus,
FA can be classified as an impulsive flare 
and FC as a long decay one. However, the
classification of flare FB is unclear since its
decay time is in the limit range between the
two types. As discussed by \citet{1988MmSAI..59...71P}
and \citet{1988A&A...201...93P},
these morphological differences may indicate
real physical differences in the energy release
process, as it also appears to happen for solar compact
and \mbox{2-ribbon} flares \citep{1977ApJ...216..108P,1981sfmh.book....1P}.
In particular, in compact flares energy is probably released only
during the impulsive phase, whereas in \mbox{2-ribbon}
flares a prolonged energy release is apparently
required to explain their long decay times.
Regarding this point, it is interesting to note the
peaks observed in the light curve during
the decay of flare FB, which are probably related to different
magnetic reconnection processes 
(superposed \mbox{flare-type} events).


\subsection{Searching for short-term variability}
\label{subsec:shortscalevariability}

There are three long intervals in the light curve of \mbox{CC Eri}
(L0, L1 and L2) that have been supposed to have a constant
emission level.
In this section
we make use of the method
given by \cite{2000A&A...353..177M} to test
whether small changes within each one of these time regions (see Fig.~\ref{fig:cceri_pn_lightcurve})
can be considered further \mbox{short-term}
stellar variability or, on the contrary, are compatible
with statistical fluctuations.

   \begin{figure}
   \centering
      \includegraphics[bb=40 571 346 808,width=8.6cm,clip]{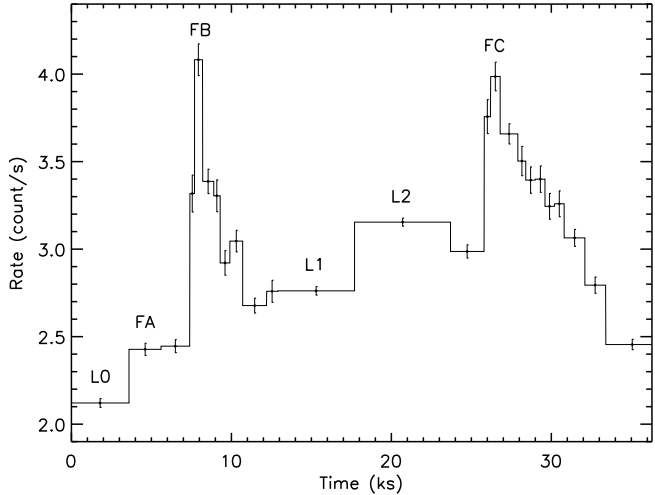}
      \caption{Rebinned light curve of CC Eri for the energy
               band \mbox{0.5 -- 10.0 keV}.
              }
         \label{fig:cceri_pn_rebinlightcurve}
   \end{figure}
%

Assuming that during the intervals L0, L1 and L2
the star has a constant \mbox{count-rate}
equal to the
mean \mbox{count-rate} characteristic of each one of these regions
($R_\mathrm{m}[\mathrm{Li}]$), we computed the net counts expected
($c_\mathrm{exp,~bin}[\mathrm{Li}]$) in the used temporal bin (\mbox{100~s}).
For every interval, we 
generated a set of simulated
data with the $N$ Poisson distribution
centered on $c_\mathrm{exp,~bin}[\mathrm{Li}]$, which represents 
possible outcomes from the observations if the
source had a constant \mbox{count-rate} equal to
the mean \mbox{count-rate} of the interval. 
Each set consisted of 1000 of such simulations.
We then calculated the
cumulative distribution function
(CDF\footnote{The cumulative distribution function
(CDF) represents the probability of observing
a number of \mbox{counts -- in the chosen temporal}
\mbox{bin -- greater} than, or equal to, a given value.}) of
the \mbox{count-rates} observed within the interval, as well as
of each one of these simulations. The set of the CDF's 
obtained for the simulations allowed us to evaluate 
the spread introduced by statistical fluctuations.

   \begin{figure}
   \centering
      \includegraphics[bb=20 57 409 333,width=8.6cm,clip]{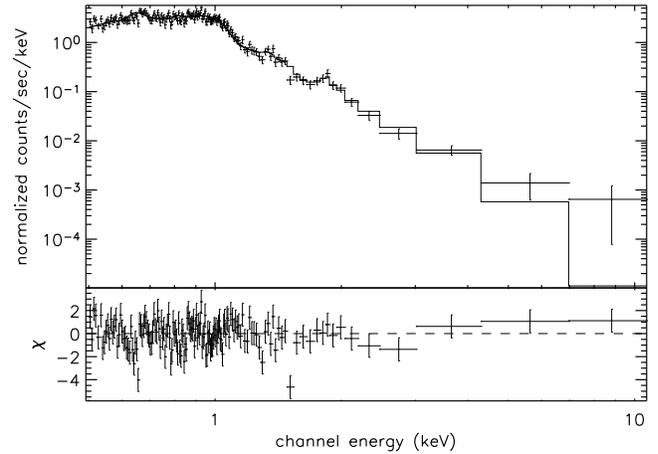}
      \caption{X-ray spectrum of the quiescent state of CC Eri (L0).
               The best fit (\mbox{2-$T$} APEC model) is also shown with
               a solid line.
              }
         \label{fig:quiescentspectrum}
   \end{figure}
%

In Fig.~\ref{fig:poissonLOL1L2} we have plotted the CDF of
the intervals L0, L1 and L2. All of them are compared with their
corresponding simulations for a constant source with
\mbox{count-rate} equal to $R_\mathrm{m}[\mathrm{Li}]$ (dashed area). 
In all the three cases the CDF of the observed \mbox{count-rates}
is contained within the space occupied by the 
CDF's of the simulations.
Therefore, statistical fluctuations can account for 
the observed spread and possible short-term 
variability on \mbox{CC Eri} cannot be distinguished from noise
in these regions. However, when the analysis is done for
the three intervals (L0, L1 and L2) altogether, 
significant differences between the observations
and simulations are found
(see Fig.~\ref{fig:poissonL0L1L2altogether}). In fact, the same
happens when the analysis is separately done for
L0 and L1, or L1 and L2. The 
difference between the emission from L0, L1 and L2 is therefore
due to stellar variability and cannot be explained
only by statistical fluctuations.

The same type of study can also be used to quantify the variations
found in the complete light curve of \mbox{CC Eri} (Fig.~\ref{fig:poissontotal}).
Since in this case we were looking for changes
relative to the quiescent emission, we chose 
as reference level the mean \mbox{count-rate} measured for the 
quiescent state, i.e. $R_\mathrm{m}[\mathrm{L0}]$. 
If the duration of the observation were long enough, the
CDF could be considered as the fraction of time that
the star spends with a \mbox{count-rate} greater than, or
equal to, a given value.
The tail 
of the CDF
at rate \mbox{$\gtrsim$ 3 count/s}
is produced by the two strongest detected flares (FB and FC). 
The rest of the CDF not consistent with
the simulations accounts for
the other kind of variability
observed 
at different time scales on the light curve of \mbox{CC Eri}.

In addition, we analyzed FA to
confirm that the emission within this \mbox{time-interval}
was not constant. We obtained that the part of the CDF
at the highest \mbox{count-rates} of FA is out of the simulations.
Therefore, FA is probably a \mbox{flare-event}.
However, the current instrumentation did not allow us
to carry out a \mbox{time-resolved} study of the different
phases of FA due to the low strength and shortness of this flare.
%

\section{Spectral analysis}
\label{sec:spectra}

In this section we carry out a time-resolved study of coronal
properties (temperature, emission measure and metal abundance)
through the whole observation of \mbox{CC Eri}.

With this scope, and taking the variability pattern in the light curve
into account, the total observing time was split in several \mbox{time-intervals} 
with approximately constant \mbox{count-rate}.
   All the intervals were chosen in order to have enough signal to
perform a reliable spectral analysis. Most of the spectra have at least
2000 counts. However, the spectrum extracted for
the rising phase of both FB and FC was taken with less
photons (\mbox{$\gtrsim$ 1000} counts) for avoiding to
mix this region with the flare maximum.
The rebinned light curve of \mbox{CC Eri}
is shown in
Fig.~\ref{fig:cceri_pn_rebinlightcurve}.
We obtained the spectra of all these intervals
following the filtering criteria
given in Sect.~\ref{sec:observations}.
Every spectrum was binned to provide
at least 8~counts per spectral bin.
Bad channels
were always excluded.

   \begin{figure*}
   \centering
      \includegraphics[bb=488 634 692 812,width=5.8cm,clip]{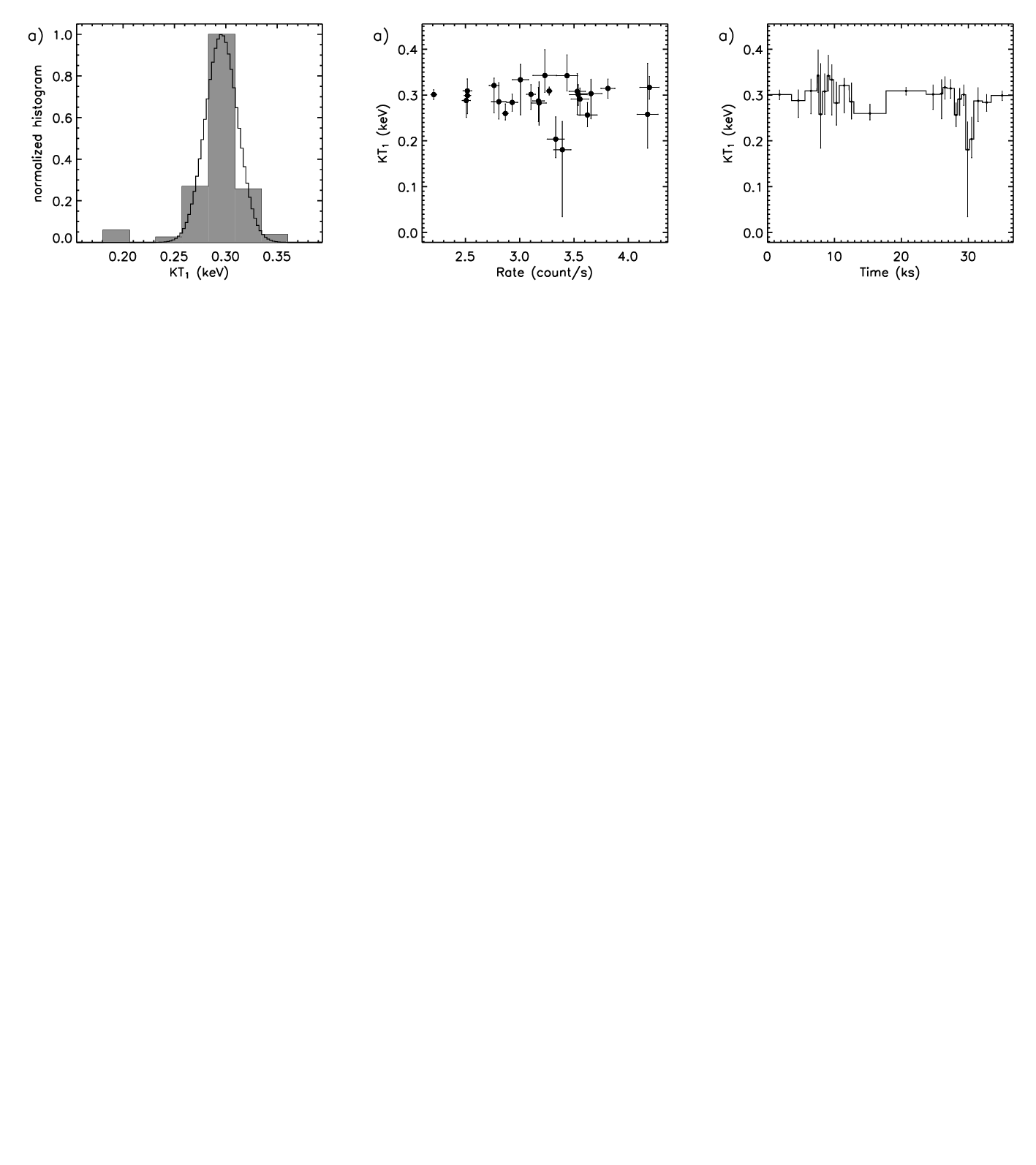}
      \includegraphics[bb=488 634 692 812,width=5.8cm,clip]{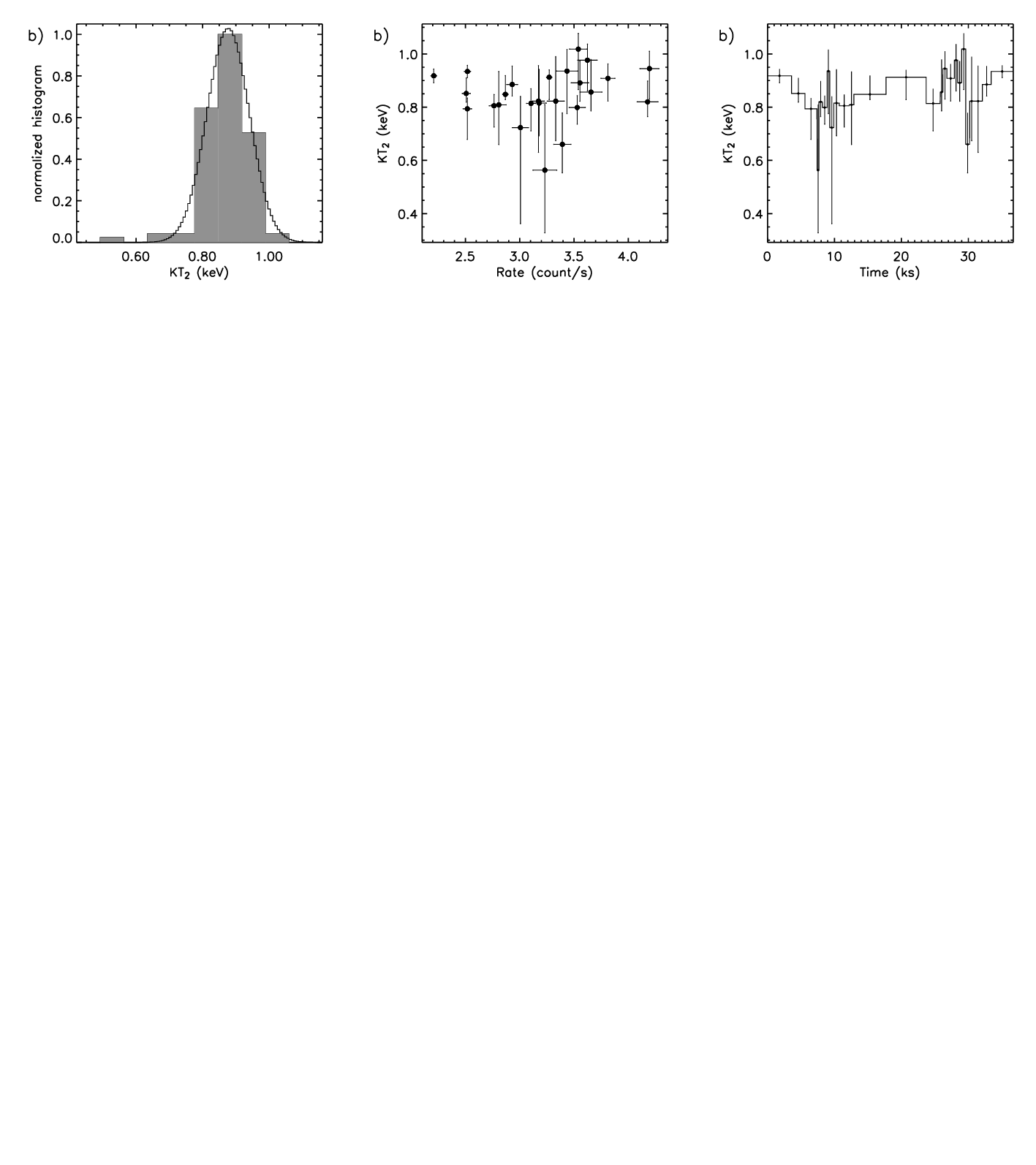}
      \includegraphics[bb=488 634 692 812,width=5.8cm,clip]{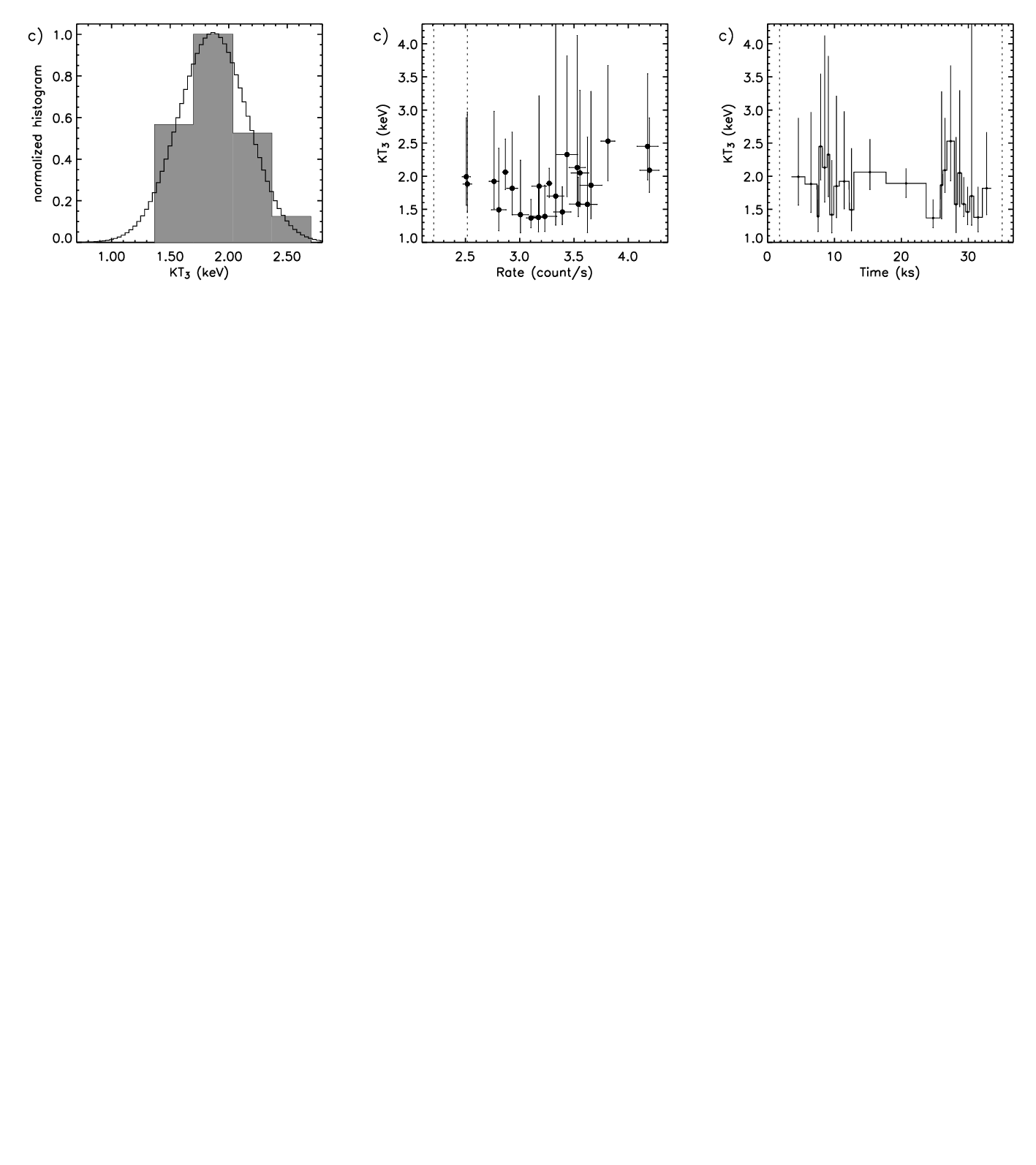}
      \caption{Time-evolution of the coronal temperatures obtained with the \mbox{3-$T$ model}.
               In the $KT_3$ panel, a dashed line marks the \mbox{time-segments} for 
               which this temperature could not be constrained (see text).
               Note that the time-resolution is the same as that shown
               in Fig.~\ref{fig:cceri_pn_rebinlightcurve}.}
         \label{fig:KTevolution}
   \end{figure*}
%

   \begin{figure*}
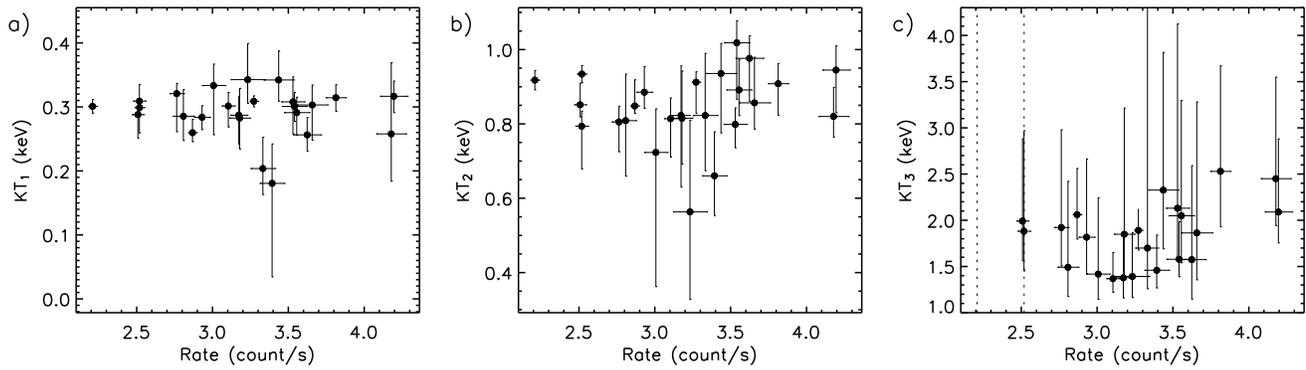

   \centering
      \includegraphics[bb=260 634 465 812,width=5.8cm,clip]{7601f07.ps}
      \includegraphics[bb=260 634 465 812,width=5.8cm,clip]{7601f08.ps}
      \includegraphics[bb=260 634 465 812,width=5.8cm,clip]{7601f09.ps}
      \caption{Coronal temperatures (obtained with the \mbox{3-$T$ model}) vs. \mbox{count-rate}.
               In the $KT_3$ panel, a dashed line marks the \mbox{time-segments} for
               which this temperature could not be constrained (see text).
               Note that no significant changes in the temperature values are observed
               even at flare maxima (points with the highest \mbox{count-rates}).
              }
         \label{fig:KTvsRate}
   \end{figure*}
%

   \begin{figure*}
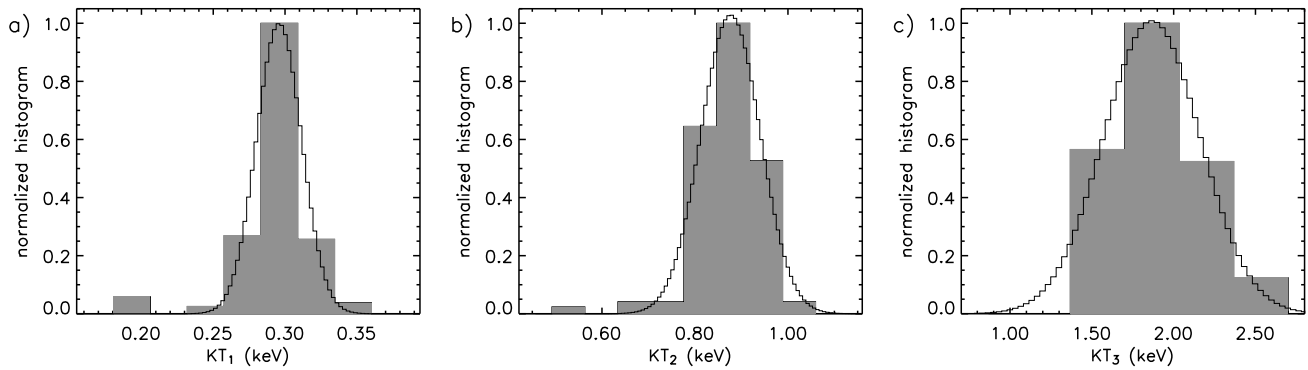

   \centering
      \includegraphics[bb=33 634 238 812,width=5.8cm,clip]{7601f07.ps}
      \includegraphics[bb=33 634 238 812,width=5.8cm,clip]{7601f08.ps}
      \includegraphics[bb=33 634 238 812,width=5.8cm,clip]{7601f09.ps}
      \caption{Distribution of the three temperatures that characterize the corona
               of \mbox{CC Eri} during the XMM-Newton observation. The corresponding
               Gaussian fits are also plotted.
              }
         \label{fig:KT_histograms}
   \end{figure*}
%

%
\begin{table*}
\caption{Results derived from the time-resolved spectral analysis
after fixing the hydrogen column density, the global coronal abundance,
and the different temperatures  
(see text in Sect.~\ref{subsec:spectral_parameters}). 
Columns show: (1) label referring to the \mbox{time-region} 
where the spectrum is included 
(see Figs.~\ref{fig:cceri_pn_lightcurve} and
\ref{fig:cceri_pn_rebinlightcurve}); (2) \mbox{time-coverage};
(3), (4) and (5) emission
measures, from the coolest ($EM_1$) to the hottest ($EM_3$) 
thermal components; (6) reduced $\chi^2$ and degrees of
freedom; (7) and (8) \mbox{X-ray} flux and luminosity in
the \mbox{0.5 -- 10.0 keV} band.}
\label{tab:emissionmeasures}
\centering
\begin{tabular}{c c c c c c c c c}     
\hline\hline
\noalign{\smallskip}
$^{(1)}$Region & $^{(2)}$Time$_\mathrm{start}$ -- Time$_\mathrm{end}$ & 
$^{(3)}$$EM_1$ & $^{(4)}$$EM_2$ & $^{(5)}$$EM_3$ & $^{(6)}$$\chi_\mathrm{red}^2$ (d.o.f) & 
$^{(7)}$$f_\mathrm{X}$\ & $^{(8)}$log $L_\mathrm{X}$(erg s$^{-1}$)\\
& (ks) & 
(10$^{52}$ cm$^{-3}$) & (10$^{52}$ cm$^{-3}$) & (10$^{52}$ cm$^{-3}$) & & (10$^{-11}$erg cm$^{-2}$ s$^{-1}$) & \\
\noalign{\smallskip}
\hline
\noalign{\smallskip}
L0 &  0.0 --  3.6 &  
1.63$_{-0.09}^{+0.09}$ & 1.43$_{-0.08}^{+0.08}$ & 0.10$_{-0.08}^{+0.08}$ & 1.44 (150) & 1.52
& 29.38\\ 
\noalign{\vskip+0.15cm}
FA &  3.6 --  5.6 &   
1.58$_{-0.13}^{+0.13}$ & 1.58$_{-0.12}^{+0.12}$ & 0.42$_{-0.11}^{+0.11}$ & 1.11 (209) & 1.77
& 29.45\\
\noalign{\vskip+0.15cm}
FB &  5.6 --  7.4 &   
1.67$_{-0.14}^{+0.14}$ & 1.54$_{-0.12}^{+0.12}$ & 0.42$_{-0.12}^{+0.12}$ & 0.97 (200) & 1.78
& 29.45\\ 
\noalign{\vskip+0.15cm}
FB &  7.4 --  7.7 &    
1.92$_{-0.37}^{+0.38}$ & 1.69$_{-0.35}^{+0.37}$ & 1.17$_{-0.39}^{+0.41}$ & 0.96 ( 85) & 2.39
& 29.58\\
\noalign{\vskip+0.15cm}
FB &  7.7 --  8.2 &   
1.55$_{-0.31}^{+0.32}$ & 2.33$_{-0.30}^{+0.31}$ & 2.05$_{-0.35}^{+0.34}$ & 0.87 (161) & 3.17
& 29.70\\
\noalign{\vskip+0.15cm}
FB &  8.2 --  8.9 &   
2.00$_{-0.25}^{+0.26}$ & 2.13$_{-0.24}^{+0.24}$ & 0.92$_{-0.24}^{+0.25}$ & 0.97 ( 98) & 2.55
& 29.61\\
\noalign{\vskip+0.15cm}
FB &  8.9 --  9.3 &   
1.60$_{-0.32}^{+0.33}$ & 1.76$_{-0.30}^{+0.31}$ & 1.65$_{-0.35}^{+0.36}$ & 1.02 (111) & 2.61
& 29.62\\
\noalign{\vskip+0.15cm}
FB &  9.3 --  9.9 &   
1.81$_{-0.25}^{+0.26}$ & 1.97$_{-0.24}^{+0.24}$ & 0.50$_{-0.23}^{+0.25}$ & 0.89 (114) & 2.14
& 29.53\\
\noalign{\vskip+0.15cm}
FB &  9.9 -- 10.7 &   
1.95$_{-0.22}^{+0.23}$ & 1.92$_{-0.21}^{+0.21}$ & 0.70$_{-0.21}^{+0.22}$ & 1.21 (166) & 2.27
& 29.56\\
\noalign{\vskip+0.15cm}
FB & 10.7 -- 12.2 &   
1.71$_{-0.15}^{+0.16}$ & 1.64$_{-0.14}^{+0.14}$ & 0.63$_{-0.14}^{+0.15}$ & 1.19 (200) & 1.98
& 29.50\\
\noalign{\vskip+0.15cm}
FB & 12.2 -- 12.9 &   
1.95$_{-0.23}^{+0.23}$ & 1.61$_{-0.21}^{+0.22}$ & 0.54$_{-0.22}^{+0.23}$ & 0.95 (147) & 1.99
& 29.50\\
\noalign{\vskip+0.15cm}
L1 & 12.9 -- 17.7 &  
1.78$_{-0.09}^{+0.09}$ & 1.67$_{-0.08}^{+0.08}$ & 0.69$_{-0.09}^{+0.09}$ & 1.22 (275) & 2.06
& 29.51\\
\noalign{\vskip+0.15cm}
L2 & 17.7 -- 23.7 &  
1.81$_{-0.08}^{+0.08}$ & 1.94$_{-0.08}^{+0.08}$ & 0.92$_{-0.09}^{+0.09}$ & 1.06 (264) & 2.37
& 29.57\\
\noalign{\vskip+0.15cm}
FC & 23.7 -- 25.8 &   
1.89$_{-0.14}^{+0.14}$ & 1.98$_{-0.13}^{+0.13}$ & 0.55$_{-0.14}^{+0.14}$ & 1.36 (234) & 2.20
& 29.54\\
\noalign{\vskip+0.15cm}
FC & 25.8 -- 26.2 &   
2.12$_{-0.33}^{+0.35}$ & 2.22$_{-0.32}^{+0.31}$ & 0.91$_{-0.32}^{+0.33}$ & 0.99 (121) & 2.64
& 29.62\\
\noalign{\vskip+0.15cm}
FC & 26.2 -- 26.8 &   
2.12$_{-0.29}^{+0.30}$ & 2.07$_{-0.27}^{+0.27}$ & 1.95$_{-0.30}^{+0.31}$ & 1.12 (175) & 3.16
& 29.70\\
\noalign{\vskip+0.15cm}
FC & 26.8 -- 27.9 &   
1.86$_{-0.21}^{+0.21}$ & 2.14$_{-0.20}^{+0.21}$ & 1.44$_{-0.23}^{+0.23}$ & 1.12 (184) & 2.81
& 29.65\\
\noalign{\vskip+0.15cm}
FC & 27.9 -- 28.4 &   
1.95$_{-0.29}^{+0.30}$ & 2.15$_{-0.28}^{+0.29}$ & 1.08$_{-0.30}^{+0.32}$ & 0.91 (142) & 2.64
& 29.62\\
\noalign{\vskip+0.15cm}
FC & 28.4 -- 29.0 &   
2.33$_{-0.27}^{+0.28}$ & 1.91$_{-0.25}^{+0.25}$ & 1.00$_{-0.26}^{+0.27}$ & 0.82 (155) & 2.57
& 29.61\\
\noalign{\vskip+0.15cm}
FC & 29.0 -- 29.6 &   
2.21$_{-0.27}^{+0.28}$ & 1.62$_{-0.25}^{+0.25}$ & 1.47$_{-0.27}^{+0.28}$ & 1.19 (160) & 2.63
& 29.62\\
\noalign{\vskip+0.15cm}
FC & 29.6 -- 30.2 &   
1.89$_{-0.27}^{+0.28}$ & 1.98$_{-0.26}^{+0.27}$ & 0.99$_{-0.27}^{+0.28}$ & 0.99 (156) & 2.47
& 29.59\\
\noalign{\vskip+0.15cm}
FC & 30.2 -- 30.8 &   
2.00$_{-0.27}^{+0.27}$ & 2.07$_{-0.25}^{+0.26}$ & 0.69$_{-0.26}^{+0.27}$ & 0.96 (149) & 2.38
& 29.58\\
\noalign{\vskip+0.15cm}
FC & 30.8 -- 32.1 &   
2.15$_{-0.19}^{+0.19}$ & 1.90$_{-0.17}^{+0.18}$ & 0.54$_{-0.17}^{+0.18}$ & 1.05 (148) & 2.24
& 29.55\\
\noalign{\vskip+0.15cm}
FC & 32.1 -- 33.4 &   
2.00$_{-0.17}^{+0.17}$ & 1.73$_{-0.15}^{+0.16}$ & 0.53$_{-0.15}^{+0.16}$ & 1.15 (190) & 2.08
& 29.52\\
\noalign{\vskip+0.15cm}
FC & 33.4 -- 36.7 &   
1.82$_{-0.10}^{+0.10}$ & 1.67$_{-0.09}^{+0.09}$ & 0.12$_{-0.09}^{+0.09}$ & 1.73 (139) & 1.74
& 29.44\\
\noalign{\smallskip}
\hline
\end{tabular}
\end{table*}
%

We modeled the spectra by using the
Astrophysical Plasma Emission Code
\citep[APEC,][]{2001ApJ...556L..91S} included in
the XSPEC software.
APEC calculates spectral models
for hot, optically thin plasmas using atomic data stored
in the Astrophysical Plasma Emission Database
\citep[APED,][]{2001ASPC..247..161S}. The APED files
contain atomic data such as
collisional and radiative rates, recombination cross
sections, dielectronic recombination rates,
and satellite line wavelengths, which constitute the
relevant information for calculating both
the continuum and line emission.
Interstellar absorption was
taken into account by using the
photoelectric cross sections
of \citet{1983ApJ...270..119M}, also
available in XSPEC.
Despite the generally satisfactory
results of using the $\chi^2$ minimization technique for obtaining the
best fitted model, it runs into problems when the number of events
is small. For this reason,
our fit procedure was based on the \mbox{Cash's {\it C} statistic} 
minimization \citep{1989ApJ...342.1207N}, which gives
better fits in the \mbox{low-count} regime (note also that the
\mbox{{\it C} statistic} is equivalent to $\chi^2$ in the limit
of large number of counts). 
Unfortunately, for the \mbox{{\it C} statistic} there exists no
method analogous to that of the reduced $\chi^2$ value ($\chi_\mathrm{red}^2$) 
with which we can measure the goodness of the fit.
We can determine the best parameters by minimizing
the function, but we have no criteria for rejecting
the model. Therefore, for giving an idea of the
goodness of our fits we show the
reduced $\chi^2$ value generated by each model (obtained
through \mbox{{\it C} statistic} minimization)
and its corresponding data set.
The errors associated with the fitted
parameters were calculated for a confidence
level of \mbox{2.706 $\sigma$}.

A 2-$T$ model was needed to describe
the shape of the coronal spectrum of the quiescent
state (L0, see Fig.~\ref{fig:quiescentspectrum}).
Leaving all the parameters as free variables, 
and taking the same global abundance \citep[scaled
on the solar photospheric values of][]{1989GeCoA..53..197A} 
for all the temperature components, we obtained the following results:
the hydrogen column density ($N_\mathrm{H}$) of the intervening
interstellar medium is negligible in the studied
wavelength range, as expected due to the proximity
of \mbox{CC Eri}; the global coronal abundance
is $Z/Z_\odot = 0.33_{-0.07}^{+0.10}$;
the temperatures are
\mbox{$KT_1 = 0.301_{-0.011}^{+0.011}$ keV} and
\mbox{$KT_2 = 0.927_{-0.036}^{+0.034}$ keV};
and the emission measures for these thermal components are
\mbox{$EM_1 = 1.69_{-0.31}^{+0.35} \times 10^{52}$} cm$^{-3}$ and
\mbox{$EM_2 = 1.50_{-0.26}^{+0.30} \times 10^{52}$} cm$^{-3}$,
that is, \mbox{$EM_2/EM_1 \approx 0.9$}.
These values give the best possible fit to the data using the {\it C}
statistic, corresponding to a 
\mbox{$\chi_\mathrm{red}^2 \approx$ 1.44}.


\subsection{Time-resolved study of spectral parameters}
\label{subsec:spectral_parameters}

We carried out the spectral analysis of all the
\mbox{time-intervals} showed in Fig.~\ref{fig:cceri_pn_rebinlightcurve}
to examine the influence of the observed flares on the
temperatures and emission measures of the plasma.
\citet{2001ApJ...557..906R} converted \mbox{X-ray} data of a sample
of solar flares into the same format and framework
as stellar \mbox{X-ray} data, in the perspective to use them
as templates for interpreting stellar flares.
They found that synthesized stellar-like spectra
of solar flares (previously subtracted by the quiescent
``background'' spectrum) are generally \mbox{well-fitted} with
a single thermal component (\mbox{1-$T$ model}) at a temperature 
close to that of the maximum of the $EM(T)$ distribution.
However, two thermal components are sometimes needed
to fit the data during the decay, probably due to a rapid
variation of the plasma temperature within this phase.
In fact, they detected deviations from the isothermal
description for the flares
with no significant sustained heating.
Consequently, our first approach consisted on using
a \mbox{3-$T$ model} to fit each observed spectrum,
fixing all the parameters of the first two components 
at their quiescent values, whereas the temperature
and emission measure of the third component
(which describes the flaring plasma)
were free to vary.
Generally, the statistics did not allow to constrain
the abundance of the additional APEC component. Therefore,
it was 
assumed to be equal to the global abundance obtained
for the quiescent state in order to avoid unphysical
solutions. Results showed that this model failed
to reproduce the observed data under the considered
assumptions. A systematic 
excess at high energies appeared in the residuals,
and the temperature of the hottest
component resulted to be very similar to that
of the second one. This suggested us that the
emission measure of the second component may
be changing (note that we fixed it at its quiescent 
value), and therefore the third component was trying to compensate
its enhancement providing a temperature lower than
that required to fit the \mbox{high-energy} region of
the spectra. For these reasons, we decided to perform
a \mbox{self-consistent} analysis by fitting the data 
using an ``iterative'' procedure, which is described below.

   \begin{figure}
   \centering
      \includegraphics[bb=22 60 352 812,width=8.6cm,clip]{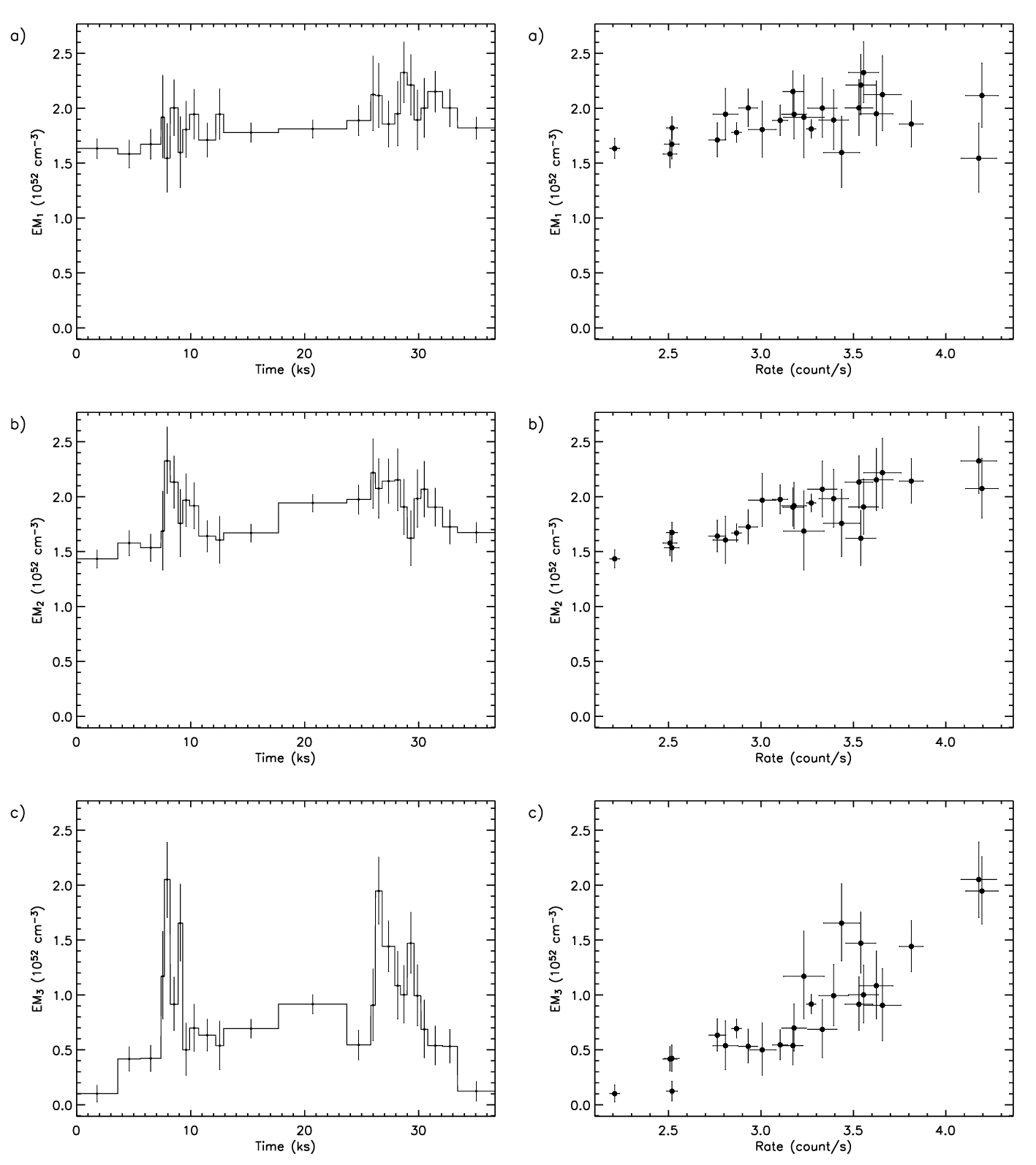}
      \caption{Time-evolution obtained for the emission measure of the three thermal
               components. Note that the time-resolution is the same as that
               shown in Fig.~\ref{fig:cceri_pn_rebinlightcurve}.
              }
         \label{fig:EMevolution}
   \end{figure}
%
   \begin{figure}
   \centering
      \includegraphics[bb=362 60 692 812,width=8.6cm,clip]{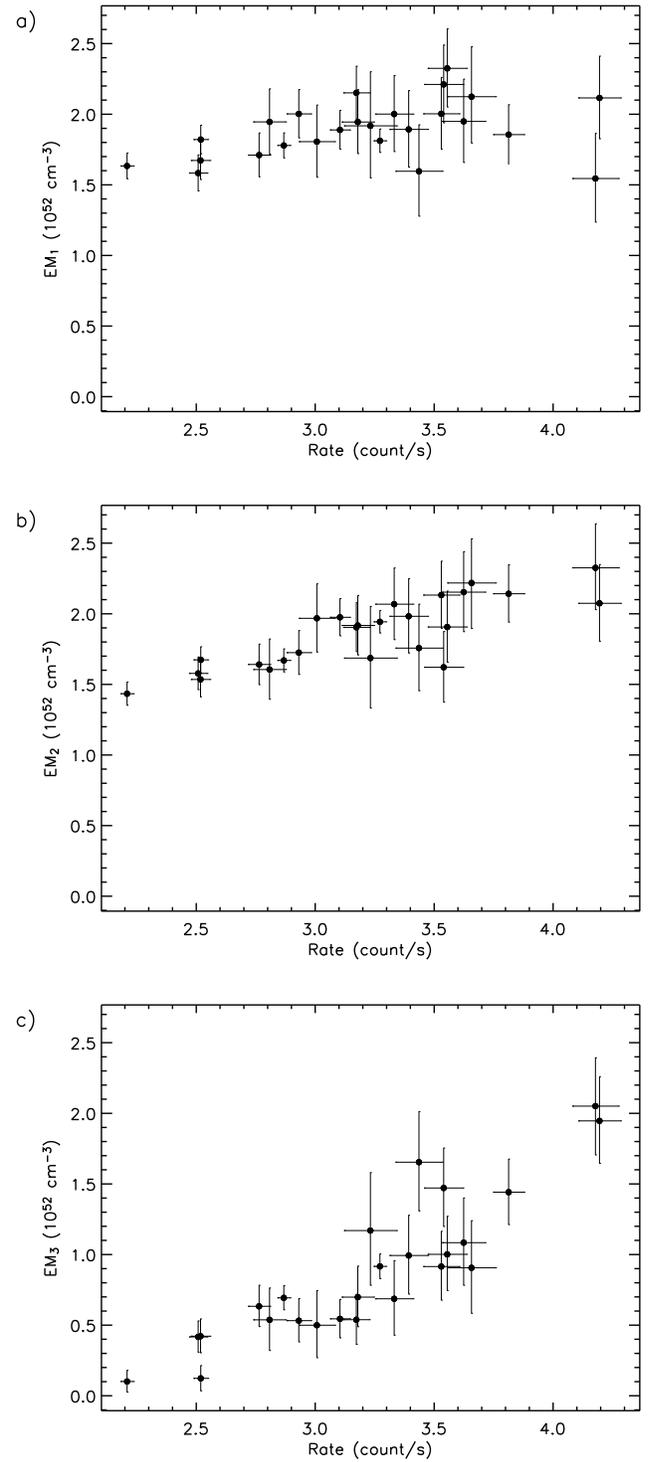}
      \caption{Emission measure of the three thermal components vs. \mbox{count-rate}.
              }
         \label{fig:EMvsRate}
   \end{figure}
%

Each spectrum was fitted with a \mbox{3-$T$ model},
taking the same global abundance for all the thermal
components and fixing it at its quiescent value.
The first iteration was done leaving all the temperatures 
and emission measures as free parameters. Results for
the temperature of the coolest component ($KT_1$) were
plotted versus time (Fig.~\ref{fig:KTevolution}a) and
\mbox{count-rate} (Fig.~\ref{fig:KTvsRate}a). No change in 
$KT_1$ was noticed.
Moreover, the normalized histogram of the \mbox{time-weighted} values
of $KT_1$ (Fig.~\ref{fig:KT_histograms}a) showed an approximately 
Gaussian distribution peaked at the temperature of the coolest
component of the \mbox{2-$T$ model} that described the quiescent state. 
A Gaussian fit provided the mean temperature of 
the distribution, together with
its uncertainty (standard deviation), that is\\

\noindent
\mbox{$\overline{KT}_1$ = 0.296 $\pm$ 0.016 keV}.\\

\noindent
This value can be considered the best estimation
for the temperature of the coolest component.
Since $\overline{KT}_1$ was constant during the observation, 
we fixed it and computed the fits again (second iteration) 
allowing variations in the other two temperatures
and in the emission measure of all the three thermal
components. No change was then detected in
$KT_2$ (see Figs.~\ref{fig:KTevolution}b and~\ref{fig:KTvsRate}b). 
This temperature also appeared to be uniform throughout the observation.
Using a Gaussian distribution to fit the normalized histogram 
of the \mbox{time-weighted} values of this parameter 
(Fig.~\ref{fig:KT_histograms}b), we obtained\\

\noindent
\mbox{$\overline{KT}_2$ = 0.88 $\pm$ 0.06 keV},\\

\noindent
which is clearly consistent with the temperature
of the hottest component of the \mbox{2-$T$ model} that 
described the quiescent state.
In addition to $\overline{KT}_1$, $\overline{KT}_2$
was also fixed in the model and the fits were
repeated (third iteration) leaving free the temperature
of the hottest component and the three emission measures.
Figs.~\ref{fig:KTevolution}c and~\ref{fig:KTvsRate}c
show the values obtained for $KT_3$
versus time and \mbox{count-rate}, respectively.
It was not possible to constrain this temperature
for the segments with an emission level equal
or very similar
to that of the quiescent state (first and last \mbox{time-intervals}).
However, we cannot discard the existence of a little amount of
hotter material during these \mbox{time-intervals}, with a contribution 
not strong enough to be detected.
No clear pattern was detected in the \mbox{time-evolution} 
of $KT_3$ during the observed flares (see Fig.~\ref{fig:KTevolution}c). 
In fact, the most surprising result is that no
significant enhancement in this temperature
was found during any phase of the observed flares
when comparing them with \mbox{non-flaring} intervals
(see also Fig.~\ref{fig:KTvsRate}c).
Furthermore, when accounting for the calculated uncertainties, all the values of
$KT_3$ are compatible with the average value obtained
by fitting a Gaussian distribution to the normalized histogram
of the \mbox{time-weighted} values of $KT_3$ (Fig.~\ref{fig:KT_histograms}c), 
as occurred for the other two temperatures, that is\\

\noindent
\mbox{$\overline{KT}_3$ = 1.87 $\pm$ 0.30 keV}.\\

\noindent
Therefore, we fixed the three temperatures ($\overline{KT}_1$, $\overline{KT}_2$
and $\overline{KT}_3$) and computed the last iteration leaving all the
emission measures as free parameters when doing the fits.
Results of the \mbox{best-fitting} models 
are given in \mbox{Table~\ref{tab:emissionmeasures}}
(note that the first and last \mbox{time-intervals}
were also fitted with the same model).
All the calculated fits can be considered \mbox{good-quality} models
in terms of the $\chi_\mathrm{red}^2$ test statistic. 
The \mbox{time-evolution}
of the three emission measures is plotted in Fig.~\ref{fig:EMevolution}.
Fig.~\ref{fig:EMvsRate} complements this figure giving
the emission measure of each component versus the \mbox{count-rate}.

Our analysis shows that the occurrence of flares may be explained with the
increasing of the emission measures related to the hottest temperatures, and in
particular to the third one, while the dominant temperatures remain
unchanged.


\subsection{Loop modeling}
\label{subsec:modeling}

Although stellar flares are not spatially resolved,
we can infer the size of these structures
by assuming that they are 
produced by the same basic physical mechanisms as 
solar flares and using 
flare loop models 
\citep[see][]{1984SoPh...93..351K,1986ApJ...301..262W,1988A&A...201...93P,1989A&A...213..245V,1990A&A...228..403P,1991A&A...241..197S,1995ApJ...453..464H,1997A&A...325..782R,2002ASPC..277..103R}.
Under the hypothesis of flares
occurring inside closed coronal structures, and
assuming that the heat pulse is released at 
the beginning of the flare, the decay time
of the X-ray emission scales with the length
of the loop which confines the flaring plasma.
However, 
the presence of significant heating during the
decay would slow down this phase and, 
therefore, the
size of the flaring loop would be overestimated. 
It was shown that the slope ($\zeta$) of the path of
the flare decay in the density-temperature
\mbox{(log $n_\mathrm{F}$ -- log $T_\mathrm{F}$)} plane mainly depends on the
heating decay time \citep{1992A&A...253..269J,1993A&A...267..586S,1997A&A...325..782R}.
\citet{1997A&A...325..782R}
derived an empirical formula using hydrodynamic simulations 
of single \mbox{semi-circular} flaring loops with constant
\mbox{cross-section}, and including the effect of the heating in the decay:
%
\begin{equation}
  L = \frac{\tau_{D}\sqrt{T_\mathrm{max}}}{3.7 \times 10^{-4} F(\zeta)}~~~~~~~~~~F(\zeta) \ge 1
 \label{eq:looplength}
 \end{equation}
where
$L$ is the loop half-length (in cm),
$T_\mathrm{max}$ the loop maximum temperature (in K),
$\tau_{D}$ the \mbox{e-folding} decay time derived from
the light curve (in seconds), and $F(\zeta)$ a \mbox{non-dimensional}
factor (larger than one) which accounts for 
the heating in the decay.
The slope of the decay path in the density-temperature
diagram is maximum (\mbox{$\sim$~2}) if the heating is negligible
during the decay, and minimum (\mbox{$\sim$~0.5}) if the heating
dominates this phase \citep{1992A&A...253..269J}.
\mbox{Eq.~(\ref{eq:looplength})} was successfully tested on resolved solar flares
observed with \mbox{Yohkoh/SXT} \citep{1997A&A...325..782R}
and has been further extended and applied to several
stellar flares observed with a wide variety of detectors
\citep{1998A&A...334.1028R,2000A&A...354.1021F,2000A&A...362..628F,2000A&A...353..987F,2001A&A...375..485F,2000A&A...356..627M,2001A&A...365L.336G,2002A&A...392..585S,2003MNRAS.345..714B,2004A&A...416..733R,2005A&A...430..155P,2006astro.ph.11632F}.
The correction factor $F(\zeta)$ needs to be
calibrated for each detector, since it depends on the spectral
bandpass and resolution. For \mbox{EPIC-PN} observations, 
the expression given by \citet{2007A&A...submitted} has to be used:
%
\begin{equation}
  F(\zeta) = \frac{c_a}{\zeta-\zeta_a} + q_a
 \label{eq:correctionfactor}
 \end{equation}
where
\begin{math}
  c_a = 0.51,~\zeta_a = 0.35,~q_a = 1.36.
 \end{math}
\mbox{Eq.~(\ref{eq:correctionfactor})} can be used for slopes in the range
\begin{math}
0.35 < \zeta \le 1.6.
 \end{math}

   \begin{figure*}
   \centering
      \includegraphics[bb=30 570 350 806.5,width=8.4cm,clip]{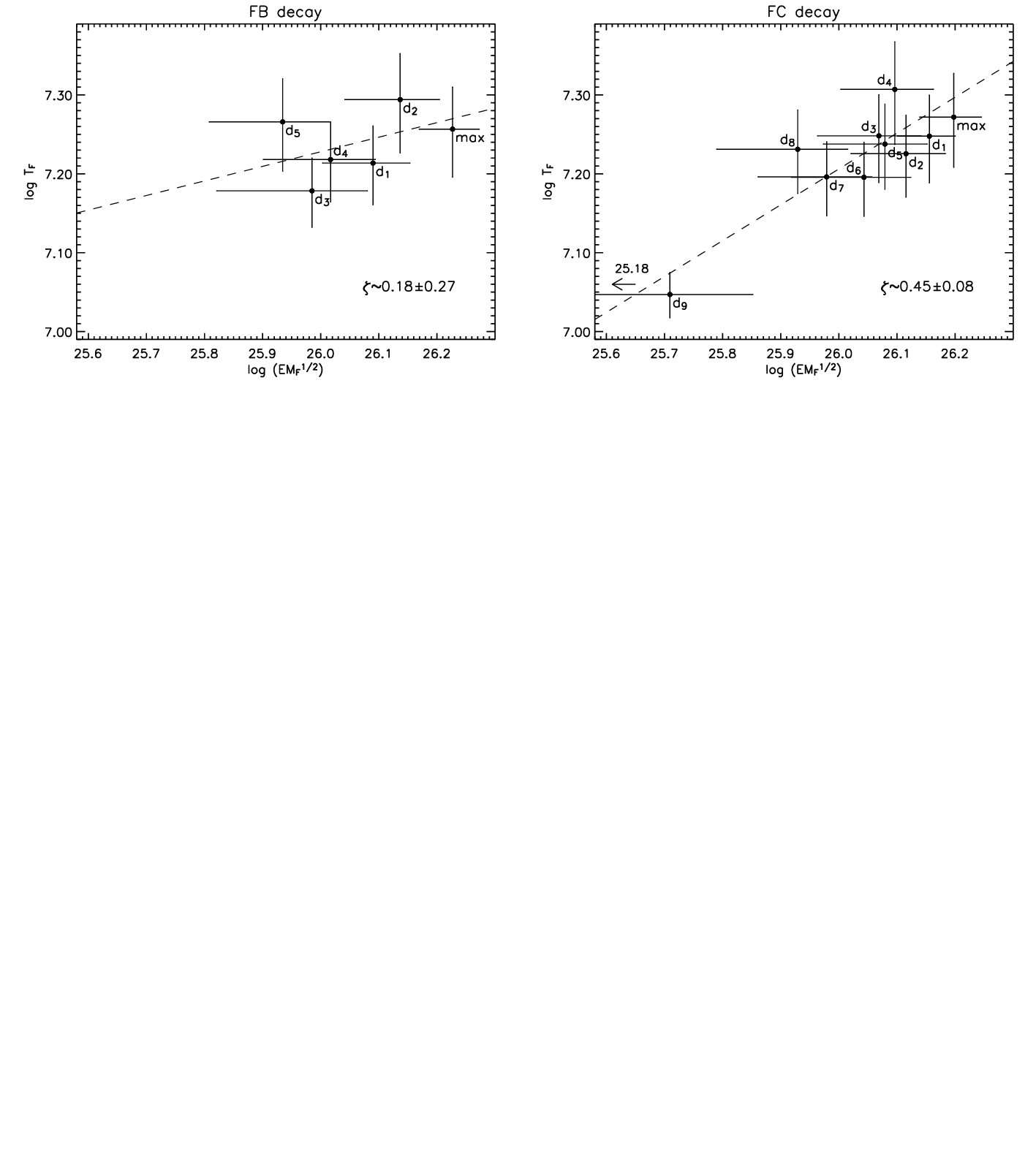}
      \includegraphics[bb=370 570 690 806.5,width=8.4cm,clip]{7601f12.ps}
      \caption{Density--temperature (log $n_\mathrm{F}$ -- log $T_\mathrm{F}$)
               diagram of the variable component (see text)
               for the decay phase of flares FB (left panel) and FC (right panel).
               $EM_\mathrm{F}^{1/2}$ has been used as a proxy for the density.
               Numbers indicate the time-evolution of the values. The point at the flare
               maximum (max) is also
               plotted. For each flare, the dashed line shows the best linear fit
               to the decay data. The derived slopes ($\zeta$) are also
               given.
              }
         \label{fig:logTvs0.5logEM}
   \end{figure*}
%

For stellar flares the density can rarely be directly measured.
Nevertheless, since reasonably the soft \mbox{X-ray} flare volume
is approximately constant during the flare development, 
the square root of the emission measure can be used as a proxy
for the density (note that 
\mbox{$EM_\mathrm{F} = \int n_\mathrm{e} n_\mathrm{H}\,dV_\mathrm{F} \approx n_\mathrm{F}^2V_\mathrm{F}$}
for a totally ionized hydrogen plasma).

As described in $\S$~\ref{subsec:spectral_parameters}, the analysed
flares (FB and FC) only involve the two hottest components of the \mbox{3-$T$ model}
used to fit the spectra. We obtained the flare contribution of each one of these components
to the total emission measure ($EM_{2,\mathrm{F}}$ and $EM_{3,\mathrm{F}}$)
by simply subtracting from $EM_{2}$ and $EM_{3}$ their corresponding value 
measured for the quiescent state (L0). Therefore, the total emission measure
of the flaring plasma ($EM_\mathrm{F}$) is given by the sum of
$EM_{2,\mathrm{F}}$ and $EM_{3,\mathrm{F}}$. A mean temperature
of the flaring plasma ($T_\mathrm{F}$) was also calculated for
each spectrum as a
weighted average of the temperature of the two hottest
components in the \mbox{3-$T$ model}, where the respective
weights were $EM_{2,\mathrm{F}}/EM_{F}$ and $EM_{3,\mathrm{F}}/EM_{F}$.
The resulting \mbox{log $EM_\mathrm{F}^{1/2}$ -- log $T_\mathrm{F}$}
diagram is plotted in Fig.~\ref{fig:logTvs0.5logEM} for the decay of FB and FC.
A linear fit to the data provided a slope of 
\mbox{$\zeta = 0.18 \pm 0.27$} for FB and \mbox{$\zeta = 0.45 \pm 0.08$ for FC},
indicating the presence of sustained heating during the decay of
both flares. Since the slope for FB is compatible with
the lower asymptotic value for which \mbox{Eq.~(\ref{eq:correctionfactor})}
can be applied, we obtained an upper limit for the loop \mbox{half-length}
of this flare by using \mbox{$\zeta = 0.45$}. 
The maximum temperature in the flaring loop ($T_\mathrm{max}$) can
be derived from the maximum observed temperature ($T_\mathrm{F,max}$).
The observed temperature is a kind of average temperature of
the flaring loop, which is therefore lower than the loop maximum temperature.  
The relationship between $T_\mathrm{max}$ and $T_\mathrm{F,max}$ 
depends on the instrumental response. For
\mbox{EPIC-PN}, \citet{2004A&A...416..733R} reported:
\begin{equation}
  T_\mathrm{max} = 0.184 T_\mathrm{F,max}^{1.130}
 \label{eq:Tmax_Tobs}
 \end{equation}
Thus, $T_\mathrm{max}$ resulted to be \mbox{$32 \times 10^{6}$ K}
for FB, and \mbox{$30 \times 10^{6}$ K} for FC, with uncertainties of
\mbox{16 $\%$}.

Using the quantities given above, the values of $\tau_{D}$ shown in
Table~\ref{tab:timeflarephases}, and \mbox{Eqs.~(\ref{eq:looplength})} 
and (\ref{eq:correctionfactor}), we obtained \mbox{$L < 7 \times 10^{9}$ cm}
for the \mbox{half-length} of the loop involved in the flare FB, and
\mbox{$L \approx (14 \pm 8) \times 10^{9}$ cm} for that involved in FC.



%

\section{Discussion and conclusions}
\label{sec:discussionconclusions}

We have presented a detailed study of the \mbox{X-ray} 
variability that the binary star \mbox{CC Eri} showed during an
\mbox{XMM-Newton} observation. The great sensitivity of the
\mbox{EPIC-PN} CCD camera allowed us to perform 
\mbox{time-resolved} spectroscopy of two flares weaker
(flux increases of factors \mbox{1.5 -- 1.9}) than those typically 
analysed in the literature. The maximum luminosity of 
these flares in the \mbox{0.5 -- 10.0 keV} band was
\mbox{$\approx 2.6 \times 10^{29}$ erg/s}.

Results show that,
during the analysed observation, the corona of \mbox{CC Eri}
is \mbox{well-described} by a \mbox{3-$T$ model}
with constant temperatures (values 
of 3, 10 and 22 MK) 
and \mbox{time-variable} emission measures. 
The emission measure of the coolest component ($EM_1$) ranges
between \mbox{1.6 -- 2.3 $\times$ 10$^{52}$ cm$^{-3}$},
the one of the second component ($EM_2$) between
\mbox{1.4 -- 2.3 $\times$ 10$^{52}$ cm$^{-3}$}, and
that of the hottest component ($EM_3$) varies
between \mbox{0.1 -- 2.1 $\times$ 10$^{52}$ cm$^{-3}$}.
That is, $EM_1$ reaches 1.4 times the value measured
for the quiescent state, $EM_2$ varies up to a factor
of 1.6, and $EM_3$ increases up to 21 times its
quiescent value. Therefore, the hottest component
dominates the X-ray variability observed on \mbox{CC Eri}
(see also Fig.~\ref{fig:EMevolution}).
The changes in the emission measure of
the coolest component could be due to the variable aspect
of the binary during its orbital motion.
However, the changes in the emission measure of the two hottest components are clearly
correlated with the variations observed in the light curve during
the detected flares (see Figs.~\ref{fig:EMevolution} and~\ref{fig:EMvsRate}).
The results obtained for the two hottest components resemble those reported by
\citet{2001ApJ...557..906R} for a sample of
solar flares that covered a wide range of intensities and
physical conditions. They found that the $EM(T)$ distribution
of all the flares in the sample follows a common evolution path:
it starts low but already at a relatively high temperature
(\mbox{$\sim$~10$^7$~K} independently of the flare intensity),
it grows toward higher $EM$ values during the rising phase
and then decreases during the decay, maintaining always a
more or less constant shape and peak temperature.
In other words, they found that the height of the
$EM(T)$ distribution is clearly variable during
the different phases of solar flares,
while the width and peak temperature of the distribution
suffer from much smaller changes.
As far as we are concerned, this is the first time that
no evident temperature variations are observed in a
\mbox{time-resolved} study of stellar flares
(see Figs.~\ref{fig:KTevolution} and~\ref{fig:KTvsRate}).
Besides, the fact that the temperatures
that characterize the flaring plasma coincide with
those of the quiescent state
is consistent with a recurrent idea in the literature, that is:
the quiescent emission of magnetically active stars
may be produced, to a large fraction,
by continuous flaring activity
(see \S~\ref{introduction} and references therein).
Thus, the light curves of multiple small flares
overlap, and only larger flares stand out from
the quiescent level. Besides, without temperature
variations, differences in the emission measure
between \mbox{time-intervals} with diverse \mbox{count-rate},
such as those observed for L0, L1 and L2
(see Fig.~\ref{fig:cceri_pn_lightcurve}
and Table~\ref{tab:emissionmeasures}),
can be identified with differences in the total coronal volume
occupied by the flaring plasma.
Density variations may also account for the changes in emission 
measure. However, we expect that, if they were significant, they would 
be driven by local heating injections or leaks, and therefore coupled to 
temperature variations but, as discussed above,
temperature variations are not detected.

The slope of the path in the \mbox{density-temperature}
diagram indicates the presence of significant heating
during the decay of the analysed flares, and leads 
to an upper limit of 7 
and \mbox{14 $\times~10^{9}$ cm}, respectively,
for the \mbox{half-length} of the flaring loops. 
From the orbital solution of \mbox{CC Eri}
\citep{2000A&A...359..159A} and the stellar radius of its components, 
we obtain a minimum distance between
their surfaces of \mbox{$\sim 1.4~\times~10^{11}$ cm}.
Even the largest loops involved in the detected flares
only cover \mbox{$\sim$ 10~$\%$} of this distance.
Therefore, it is likely that the flaring loops are included all in
the corona of one of the stars, and do not
result from the magnetic fields bridging the two stars.
In addition, bearing the spectral type of the stellar 
components of \mbox{CC Eri} in mind, the ratio between 
the \mbox{half-length} of the flaring loops and the stellar radius ($R_\mathrm{*}$)
would be 0.1 and 0.3 if the flares were produced by the
K7.5Ve star, or 0.2 and 0.4 if they were produced by the
M3.5Ve star. In both cases, the loop height above the stellar 
surface ($2L/\pi$ for vertical loops with \mbox{semi-circular} 
geometry) is about \mbox{0.1 -- 0.3 $R_\mathrm{*}$}, implying a relatively 
compact flaring corona. These results are in
agreement with the loop \mbox{half-lengths} derived 
for other Me dwarfs \citep[$L \le 0.5~R_\mathrm{*}$;][]{2003SSRv..108..577F}.

%

\begin{table}
\begin{minipage}[t]{\columnwidth}
\caption{Parameters derived for flares FB and FC.}
\label{tab:flareparameters}
\centering
\renewcommand{\footnoterule}{}  
\begin{tabular}{l l l}

\hline\hline
\noalign{\smallskip}
 & Flare FB & Flare FC \\
\noalign{\smallskip}
\hline
\noalign{\smallskip}
\vspace{0.1cm}
\footnote{e-folding decay time derived from the light curve.}$\tau_\mathrm{D}$ ($\mathrm{s}$)                     
     & 2850            & 5960 \\
\vspace{0.1cm}
\footnote{Luminosity at the flare peak (\mbox{0.5 -- 10.0 keV} band).}$L_\mathrm{X, max}$ ($10^{29}~\mathrm{erg~s^{-1}}$)  
     & 2.6             & 2.6 \\
\vspace{0.1cm}
\footnote{Total radiated energy (\mbox{0.5 -- 10.0 keV} band).}$E_\mathrm{X, tot}$ ($10^{33}~\mathrm{erg}$)         
     & 0.75             & 1.5 \\
\vspace{0.1cm}
\footnote{Slope of the decay path in the density-temperature diagram (see details in $\S$~\ref{subsec:modeling}).}$\zeta$                                              
     & 0.18 $\pm$ 0.27 & 0.45 $\pm$ 0.08 \\
\vspace{0.1cm}
\footnote{Maximum temperature in the loop at the flare peak (see $\S$~\ref{subsec:modeling}).}$T_\mathrm{max}$ ($10^{6}~\mathrm{K}$)               
     & 32 $\pm$ 5      & 30 $\pm$ 5 \\
\vspace{0.1cm}
\footnote{Half-length of the flaring loops (see $\S$~\ref{subsec:modeling}).}$L$ ($10^{9}~\mathrm{cm}$)                           
     & $<$ 7           & 14 $\pm$ 8 \\
\vspace{0.1cm}
\footnote{Maximum pressure in the loop at the flare peak \citep[estimated from the loop scaling laws given by][]{1978ApJ...220..643R}.}$p$ ($10^{3}~\mathrm{dyn~cm^{-2}}$)                  
     & $>$ 1.7          & 0.74 $\pm$ 0.34 \\
\vspace{0.1cm}
\footnote{Maximum electron density in the loop at the flare peak. We have assumed a totally ionized hydrogen plasma (i.e., $p = 2 n_\mathrm{e} k T_\mathrm{max}$).}$n_\mathrm{e}$ ($10^{11}~\mathrm{cm^{-3}}$)          
     & $>$ 2.0          & 0.88 $\pm$ 0.43 \\
\vspace{0.1cm}
\footnote{Volume of the flaring region (note that \mbox{$EM_\mathrm{F} \approx n_\mathrm{e}^2V_\mathrm{F}$} for a totally ionized hydrogen plasma).}$V_\mathrm{F}$ ($10^{30}~\mathrm{cm^{3}}$)                      
     & $<$ 0.7           & 3.2 \\
\vspace{0.1cm}
\footnote{Heating rate per unit volume at the flare peak \citep[estimated from the loop scaling laws given by][]{1978ApJ...220..643R}.}$E_\mathrm{H}$ ($\mathrm{erg~s^{-1}~cm^{-3}}$)     
     & $>$ 3.7 & $\sim$ 0.8 \\
\vspace{0.1cm}
\footnote{Number of loops needed to fill the flare volume. We have assumed a loop aspect $r/L = 0.1$ for a single loop.}$N_\mathrm{loops}$                                  
     & $\sim$ 38       & $\sim$ 19 \\
\vspace{0.1cm}
\footnote{Magnetic field ($p \le p_{B} = B^{2}/8\pi$).}$B$ ($\mathrm{G}$)                               
     & $>$ 210         & $>$ 140 \\

\noalign{\smallskip}
\hline
\end{tabular}
\end{minipage}
\end{table}
%

Assuming that at the flare peak the flaring loop is not far
from a \mbox{steady-state} condition, and given that the
derived loop \mbox{half-lengths} are significantly smaller
than the pressure scale height\footnote{The pressure scale 
height is defined as $h_\mathrm{p} = 2kT_\mathrm{max}/(\mu g)$, 
where $\mu$ is the average atomic weight and $g$ is the surface
gravity of the star. Therefore,
$h_\mathrm{p} \approx 5000 T_\mathrm{max}/(g/g_\mathrm{\odot})$.
Taking the spectral type of the stellar components of \mbox{CC Eri}
into account, \mbox{$h_\mathrm{p} \ge 7 \times 10^{10}$ cm} for both FB and FC.},
we can apply the \mbox{so-called} RTV scaling laws \citep{1978ApJ...220..643R}. 
These relationships link the pressure ($p$) and the heating rate per unit volume 
($E_\mathrm{H}$) with the loop \mbox{half-length} and the loop maximum 
temperature. Table~\ref{tab:flareparameters} lists all these quantities
for the observed flares. We have also estimated the electron density 
under the assumption of a totally ionized hydrogen plasma,
obtaining $n_\mathrm{e} \sim 10^{11}$ cm$^{-3}$. This is 
compatible with values expected for a plasma in coronal conditions.
In order for the electron density to be consistent with the
$EM_\mathrm{F}$ measured at the peak of the analysed flares,
the flaring volumes should be as large as \mbox{$10^{30}$ cm$^{3}$}.
From the pressure of the flaring plasma (see Table~\ref{tab:flareparameters}), 
we infer that the minimum magnetic field required to confine 
the plasma at the flare peak is \mbox{$\sim$ 210 G} for FB
and \mbox{$\sim$ 140 G} for FC. 

To satisfy the energy balance relation
for the flaring region as a whole, the maximum X-ray luminosity
must be lower than the total input energy rate at the flare peak
($H = E_\mathrm{H} V_\mathrm{F}$). The rest of
the input energy is used for thermal conduction, kinetic energy
and radiation at lower frequencies.
From the analysis of a large optical flare on the M3.5Ve star \mbox{AD Leo},
\citet{1993A&A...278..109H} concluded that the total kinetic energy during the event 
was of the same order as the radiated energy.
For both FB and FC, we obtain that $L_\mathrm{X, max}$ is about
\mbox{$10~\%$} of $H$, compatible with the \mbox{X-ray} radiation
being only one of the energy loss terms during the detected flares.
This value is in agreement with those reported for solar
flares, where the soft \mbox{X-ray} radiation at the peak
only accounts for \mbox{10 -- 20~$\%$} of the total
energy budget \citep{1986NASAConfPub..2439..5}.
\citet{2004A&A...416..733R} found
a similar percentage (\mbox{$\sim 15~\%$}) for a flare 
observed on the M5.5Ve star \mbox{Proxima Centauri},
while \citet{2000A&A...353..987F} estimated a higher value (\mbox{$\sim 35~\%$}) for
an extreme X-ray flare detected on the M3.5Ve star \mbox{EV Lac}.

If the detected flares were produced by a single loop, its
aspect ($\beta = r/L$, where $r$ is the radius of the loop 
\mbox{cross-section} derived from the volume and loop length) 
should be 0.6 for FB and 0.4 for FC.
Such a large \mbox{cross-section} is not observed
on solar coronal loops, for which typical values of
$\beta$ are in the range \mbox{0.1 -- 0.3}. 
Therefore, we suggest a more realistic scenario
consisting on flaring structures made up of several
similar loops. Assuming $\beta = 0.1$, 
FB and FC occur in arcades composed of $\sim$ 38 and 19
loops, respectively. Similar structures are also
observed to flare on the Sun. For example, the 
Bastille day flare (2000 July 14) is an intense
solar flare (GOES class X6) that occurred on a 
curved arcade with some 100 post-flare loops 
\citep{2001SoPh..204...91A,2002ApJ...578..590R}.
Stellar analogues have been proposed for the
Me dwarf \mbox{Proxima Centauri}, where the
best description for a flare analysed by 
\citet{2004A&A...416..733R} shows the presence of 
an arcade made up of \mbox{$\sim$ 5} loops; 
and also for the younger 
($\sim 100$ Myr) G9 dwarf \mbox{ZS 76}, where \citet{2005A&A...430..155P}
estimated \mbox{20 -- 30} loops for the flaring arcades.
Therefore, we conclude that events like solar arcade flares
may be a common phenomenon on stars, in wide generality.


\begin{acknowledgements}
   I.C.C. acknowledges support from the Spanish Ministerio de 
   Educaci\'on y Ciencia (MEC), under the grant \mbox{AP2001-0475}
   (programa nacional de Formaci\'on de Profesorado Universitario)
   and project \mbox{AYA2005-02750} (Programa Nacional de Astronom\'{\i}a y 
   Astrof\'{\i}sica).
   The work of J.L.S. has been supported by the Marie Curie Fellowship
   Contract No. MTKD-CT-2004-002769.
   We thank John Pye for useful comments which have contributed 
   to improve the manuscript.

\end{acknowledgements}

\end{document}